\documentclass[11pt,letterpaper]{JHEP3}
\usepackage{cite}
\usepackage{epsfig}
\usepackage{amsfonts}

\def \beq {\begin{equation}}
\def \eeq {\end{equation}}
\def \bea {\begin{eqnarray}}
\def \eea {\end{eqnarray}}

 \def\e{{\rm e}} 
\def\dd{{\rm d}}   
  
\def\Z#1{_{\lower2pt\hbox{$\scriptstyle#1$}}}
\def\X#1{_{\lower2pt\hbox{$\scriptscriptstyle#1$}}}

\title{\bf Gauss-Bonnet cosmologies: crossing the phantom divide
and the transition from matter
dominance to dark energy}

\author{Ben M. Leith\\
Department of Physics and Astronomy, University of Canterbury\\
Private Bag 4800, Christchurch 8020, New Zealand
\email{E-mail:b.leith@phys.canterbury.ac.nz}}
\author{Ishwaree P. Neupane\\
Department of Physics and Astronomy, University of Canterbury\\
Private Bag 4800, Christchurch 8020, New Zealand \\
and \\Theory Division, CERN, CH-1211 Geneva 23, Switzerland
\email{E-mail:ishwaree.neupane@canterbury.ac.nz}}

\abstract{ Dark energy cosmologies with an equation of state
parameter $w$ less than $-1$ are often found to violate the null
energy condition and show unstable behaviour. A solution to this
problem may require the existence of a consistent effective theory
that violates the null energy condition only momentarily and does
not develop any instabilities or other pathological features for a
late time cosmology. A model which incorporates a dynamical scalar
field $\varphi$ coupled to the quadratic Riemann invariant of the
Gauss-Bonnet form is a viable proposal. Such an effective theory
is shown to admit nonsingular cosmological evolutions for a wide
range of scalar-Gauss-Bonnet coupling. We discuss the conditions
for which our model yields observationally supported spectra of
scalar and tensor fluctuations, under cosmological perturbations.
The model can provide a reasonable explanation for the transition
from matter dominance to dark energy regime and the late time
cosmic acceleration, offering an interesting testing ground for
investigations of the cosmological modified gravity. }

\preprint{CERN-PH-TH-2006-273, \quad \hepth{0702002}}

\begin{document}

\section{Introduction and Overview}

Dark energy and its associated cosmic acceleration problem has
been presented as three well posed conundrums in modern cosmology:
(1) why the dark energy equation of state $w\Z{\rm DE}$ is so
close to $-1$, (2) why the dark energy density is comparable to
the matter density {\it now}, the so-called cosmic coincidence
problem, and (3) why the cosmological vacuum energy is small and
positive. Additionally, we do not yet have a good theoretical
understanding of the result that the expansion of our universe is
accelerating after a long period of deceleration as inferred from
the observations of type Ia supernovae, gravitational weak lensing
and cosmic microwave background (CMB) anisotropies
~\cite{Riess:2004nr,Spergel:2003cb}; for review, see, e.g.
\cite{Padmanabhan02ji}.

A possible source of the late time acceleration is given by vacuum
energy with a constant equation of state, known as the {\it
cosmological constant}, or a variant of it, {\it
quintessence}~\cite{Quintessence}. Several plausible cosmological
scenarios have also been put forward to explain these problems
within the context of modified theories of scalar-tensor gravity
such as, coupled quintessence~\cite{Amendola99a}, k-inflation or
dilatonic-ghost model~\cite{Picon-et-al}, scalar-phantom
model~\cite{Starobinsky00a}, Gauss-Bonnet dark
energy~\cite{Nojiri05b} and its various
generalizations~\cite{follow-up,IPN05d,IPN06b}. These ideas are
interesting as, like quintessence, they offer a possible solution
to the cosmic coincidence problem. The proposals in
\cite{Starobinsky00a,Nojiri05b,follow-up,IPN05d,IPN06b} are
promising because they may lead to the observationally supported
equation of state, $w\approx -1$, while more interestingly provide
a natural link between cosmic acceleration and fundamental
particle theories, such as superstring theory. In this paper we
will focus primarily on the first two questions raised above
within the context of a generalized theory of scalar-tensor
gravity that includes the Gauss-Bonnet curvature invariant coupled
to a scalar field $\varphi$.

The question of whether the gravitational vacuum energy is
something other than a pure cosmological term will not be central
to our discussion. But we note that Einstein's general relativity
supplemented with a cosmological constant term does not appear to
have any advantages over scalar-tensor gravity containing a
standard scalar potential. The recent observation that the dark
energy equation of state parameter $w$ is $\approx -1$ does not
necessarily imply that the dark energy is in the form of a
cosmological constant; it is quite plausible that after inflation
the scalar field $\varphi$ has almost frozen, so that
$V(\phi)\approx {\rm const} = \Lambda$. In a cosmological
background, there is no deep reason for expecting the energy
density of the gravitational vacuum to be a constant, instead
perhaps it can be determined by the underlying theory, as in the
case where a scalar potential possesses many minima. It is thus
worth exploring dynamical dark energy models, supporting both
$w<-1$ and $w>-1$, and also $w^{*}(z)\equiv dw/dz \ne 0$ (where
$z$ is the redshift parameter).

The expansion of the universe is perhaps best described by a
monotonically decreasing Hubble expansion rate, implying that
$\dot{H} \equiv \ddot{a}/a-\dot{a}^2/a^2 \le 0$, where $a=a(t)$ is
the scale factor of a four-dimensional Friedmann-Robertson-Walker
universe and the dot represents a derivative with respect to the
cosmic time $t$. In the presence of a barotropic fluid of pressure
$p$ and energy density $\rho$, this last inequality implies that
the cosmic expansion obeys the null energy condition (NEC),
$p+\rho\ge 0$, and hence $w\equiv p/\rho\ge -1$. The standard view
is that the Hubble expansion rate increases as we consider earlier
epochs until it is of a similar order to that of the Planck mass,
$m_{Pl}\sim {10}^{18}\,GeV$. However, in strong gravitational
fields, such as, during inflation, the Einstein description of
gravity is thought to break down and quantum gravity effects are
expected to become important. This provides a basis for the
assumption that the expansion of the universe is inseparable from
the issue of the ultraviolet completion of gravity. In recent
years, several proposals have been made in order to establish such
a link. For instance, reference~\cite{Creminelli06a} introduced a
system of a derivatively coupled scalar Lagrangian which violates
the condition $\dot{H}\le 0$ spontaneously: the model would
involve a short-scale (quantum) instability associated with a
super-luminal cosmic expansion (see also~\cite{Aref'eva:2006a}).

The beauty of Einstein's theory is in its simplicity. It has been
remarkably successful as a classical theory of gravitational
interactions from scales of millimeters through to kiloparsecs.
Thus any modification of Einstein's theory, both at small and
large distance scales, must be consistent with known tests.
Several proposals in the literature, e.g.
in~\cite{Creminelli06a,Carroll:2003wy}, do not seem to fall into
this category as these ideas would involve modifications of
Einstein's theory in a rather non-standard (and non-trivial) way.

We motivate our work through the theoretical insights of
superstring or M-theory as it would appear worthwhile to explore
the cosmological implications of such models. In this paper we
examine whether we can achieve observationally supported
cosmological perturbations in the low-energy string effective
action, which includes a non-trivial interaction between a
dynamical scalar field $\varphi$ and a Riemann invariant of the
Gauss-Bonnet form, and study its phenomenological viability as a
dark energy model. It is appreciated that a generalized theory of
scalar-tensor gravity, featuring one or several scalar fields
coupled to a spacetime curvature, or a Riemann curvature
invariant, can easily account for an accelerated universe with
quintessence ($w>-1$), cosmological constant ($w=-1$) or phantom
($w<-1$) equation of state without introducing the wrong sign on
the scalar kinetic term.

This paper is organised as follows. In the following section we
discuss a general scalar-field Lagrangian framework and write
equations of motion that describe gravity and a scalar field
$\varphi$, allowing non-trivial matter-scalar couplings. We
discuss some astrophysical and cosmological constraints on the
model. In section 3 we present the construction of a number of
scalar potentials for some specific constraints, in an attempt to
gain insight into the behaviour of the scalar potential for late
time cosmology. In section 4 we discuss inflationary cosmologies
for specific cases and study the parameters related to
cosmological perturbations of the background solution. The problem
of suitable initial conditions, given stable observationally
viable solutions with a full array of background fields for the
general system is considered in section 5 using numerical and
analytic techniques for both minimal and non-minimal scalar-matter
couplings. In section 6 we present several remarks about the
existence of superluminal propagation and/or short-scale negative
instabilities for the tensor modes. Section 7 is devoted to the
discussions of our main results.

\section{Essential Ingredients}

An unambiguous and natural way of modifying Einstein's General
Relativity in four dimensions is to introduce one or more
fundamental scalar fields and their interactions with the leading
order curvature terms in the $\alpha^\prime$ expansion, as arising
in string or M theory compactifications from ten- or
eleven-dimensions to four-dimensions. The simplest version of such
scalar-tensor gravities, which is sufficiently general for
explaining the cosmological evolution of our universe, may be
given by
\begin{equation}\label{effective}
S= S_{grav} + S_{m},
\end{equation}
with the following general action
\begin{eqnarray}
S_{grav} &=& \int d^{4}{x}\sqrt{-g}\left (\frac{R}{2 \kappa^2}
-\frac{\gamma}{2} \, (\nabla\varphi)^2-
V(\varphi) - \frac{1}{8} f(\varphi) {\cal R}_{GB}^2\right),\label{gravi-eq}\\
S_{m} &=& S (\varphi, A^2(\varphi) g_{\mu\nu},\psi_m) =\int d^4 x
\sqrt{-g} \left(A^4(\varphi) \left(\rho_m + \rho_r\right)\right),
\label{matter-eq}
\end{eqnarray}
where ${{\cal R}_{GB}^2} \equiv R^2-4 \, R_{\mu\nu} \, R^{\mu\nu}
+ R_{\mu\nu\rho\sigma} \, R^{\mu\nu\rho\sigma}$ is the
Gauss-Bonnet (GB) curvature invariant. We shall assume that
$\varphi$ is a canonical field, so $\gamma>0 $. The coupling
$f(\varphi)$ between $\varphi$ and the GB term is a universal
feature of all four-dimensional manifestations of heterotic
superstring and M theory~\cite{Antoniadis93a,Heterotic-and-M}. For
example, such a form arises at heterotic string tree-level if
$\varphi$ represents a dilaton, and at one-loop level if $\varphi$
represents the average volume modulus; in a known example of
heterotic string theory, one has $f(\varphi) \propto \sum_{n=1}
c_n\, e^{(n-2)\varphi}>0$ in the former case, while
$f(\varphi)\propto \varphi-\frac{\pi}{3}\,e^{\varphi}+4\sum_{n=1}
\ln\left(1-e^{-2n\pi\,e^\varphi}\right)+\ln 2<0$ in the latter
case (see also discussions in ~\cite{Hererotic-FollowUp}). As
discussed in~\cite{IPN05d,IPN06b}, a non-trivial or non-constant
$f(\varphi)$ is useful not only for modelling a late time
cosmology but also desirable for embedding the model in a
fundamental theory, such as superstring theory. Here we also note
that from a model building point of view, only two of the
functions $V(\varphi), f(\varphi)$ and $A(\varphi)$ are
independent; these can be related through the Klein-Gordan
equation (see below).

One can supplement the above action with other higher derivative
terms, such as those proportional to $f(\varphi)
(\nabla_\lambda\varphi \nabla^\lambda\varphi)^2$ and curvature
terms, but in such cases it would only be possible to get
approximate (asymptotic) solutions, so we limit ourselves to the
above action. In the model previously studied by Antoniadis et
al.~\cite{Antoniadis93a}, $V(\varphi)$ and $S_m$ were set to zero.
However, the states of string or M theory are known to include
extended objects of various dimensionalities, known as `branes,
beside trapped fluxes and non-trivial cycles or geometries in the
internal (Calabi-Yau) spaces. It is also natural to expect a
non-vanishing potential to arise in the four-dimensional string
theory action due to some non-pertubative effects of branes and
fluxes. With supersymmetry broken, such a potential can have
isolated minima with massive scalars, this then avoids the problem
with runway behaviour of $\varphi$ after inflation.

In this paper we fully extend the earlier analysis of the model
in~\cite{IPN05d} by introducing background matter and radiation
fields. We allow $\varphi$ to couple with both an ordinary
dust-like matter and a relativistic fluid. We show that the model
under consideration is sufficient to solve all major cosmological
conundrums of concordance cosmology, notably the transition from
matter dominance to a dark energy regime and the late time cosmic
acceleration problem attributed to dark energy, satisfying $w_{\rm
eff}\approx -1$.

\subsection{Basic equations}

In order to analyse the model we take a four-dimensional spacetime
metric in standard Friedmann-Robertson-Walker (FRW) form:
$ds^2=-dt^2 + a^2(t) \sum_{i=1}^3 (dx^i)^2$, where $a(t)$ is the
scale factor of the universe. The equations of motion that
describe gravity, the scalar field $\varphi$, and the background
fluid (matter and radiation) are given by
\begin{eqnarray}
- \frac{3}{\kappa^2} + 3 \dot{\varphi} H f_{,\,\varphi}  +
\frac{\gamma}{2} x^2 + {y} + 3
\Omega_b &=& 0, \label{GB1} \\
\frac{1}{\kappa^2} \left( 2\epsilon+3\right) + \frac{\gamma}{2}
x^2 - y -\ddot{f} - 2\dot{\varphi} H f_{,\,\varphi}(1+\epsilon)
+ 3 w_b \Omega_b &=& 0, \label{GB2}\\
\gamma\left(\ddot{\varphi} + 3 H \, \dot{\varphi}\right) +
V_{,\,\varphi}+ {3} \left(1+\epsilon\right) f_{,\,\varphi} H^4 -
\eta Q A^4 \rho_m &=& 0, \label{GB3}
\end{eqnarray} where $\ddot{f}\equiv f_{,\,\varphi\varphi}\,\dot{\varphi}^2+
f_{,\,\varphi}\,\ddot{\varphi}$, $V_{,\,\varphi}\equiv
dV/d\varphi$, $\eta$ is a numerical parameter which we define
below and
\begin{eqnarray}
&& \epsilon= \frac{\dot{H}}{H^2}, \quad {x} =
\frac{\dot\varphi}{H}, \quad y = \frac{V(\varphi)}{H^2},\quad
w_b\equiv \frac{p_b}{\rho_b}, \quad \Omega_b \equiv \frac{\rho_b
A^4}{3H^2}, \quad Q\equiv \frac{d\ln A(\varphi)}{d\varphi},
\end{eqnarray}
where $b$ stands for the background matter and radiation. Let us
also define
\begin{equation}\label{other-defs}
 \Omega_\varphi\equiv
\frac{\rho_\varphi}{3H^2}=\frac{\gamma x^2+ 2y}{6}, \qquad u\equiv
f_{,\,\varphi} H^2, \qquad \Omega_{GB}= \dot{\varphi} H
f_{,\,\varphi}=u x \equiv \mu,
\end{equation}
so that the constraint equation (\ref{GB1}) reads
$\Omega_\varphi+\Omega_{GB}+\Omega_b=1$. The density fraction
$\Omega_b$ may be split into radiation and matter components:
$\Omega_b= \Omega_r +\Omega_m$ and $w_b \Omega_b=w_m \Omega_m +
w_r \Omega_r$. Similarly, in the component form, $Q \rho_m= Q_i
\rho_m^{(i)}$. Stiff matter for which $w_m=1$ may also be
included. The analysis of Steinhardt and
Turok~\cite{Steinhardt-Turok1} neglects such a contribution, where
only the ordinary pressureless dust ($w_m=0$) and radiation
($w_r=1/3$) were considered, in a model with $f(\varphi)=0$. The
implicit assumption above is that matter couples to $A^2(\varphi)
g_{\mu\nu}$ with scale factor $\hat{a}$, where $\hat{a}\equiv a
A(\varphi)$, rather than the Einstein metric $g_{\mu\nu}$ alone,
and $\rho_r\propto 1/\hat{a}^{4}$, so $\rho_r$ does not enter the
$\varphi$ equation of motion, (\ref{GB3}). That is, by
construction, the coupling of $\varphi$ to radiation is vanishing.
This is, in fact, consistent with the fact that the quantity $Q$
couples to the trace of the matter stress tensor,
$g_{(i)}^{\mu\nu} T_{\mu\nu}^{(i)}$, which is vanishing for the
radiation component, $T_\mu^\mu=-\rho\Z{r}+3P\Z{r}=0$. We also
note that, in general, $\rho \propto 1/\hat{a}^{3(1+w)}$, thus,
for ordinary matter or dust ($w=0$), we have $\eta=1$, while for a
highly relativistic matter ($w=1$), we have $\eta =-2$.
Especially, for radiation ($w=1/3$), we have $\eta=0$; while, for
any other relativistic matter having a pressure $\rho> p
> \rho/3$, $-2< \eta < 0$.

Equations (\ref{GB1})-(\ref{GB3}) may be supplemented with the
equation of motion for a barotropic perfect fluid, which is given
by
\begin{equation}\label{conservation-eq}
\hat{a}
\frac{d\rho_b}{d\hat{a}}=\frac{1}{H}\frac{\partial\rho_b}{\partial
t}+\frac{\dot{\varphi}}{H}
\frac{\partial\rho_b}{\partial\varphi}=-3(1+w_b)\rho_b,
\end{equation}
where $p_b$ is the pressure of the fluid component with energy
density $\rho_b$. In the case of minimal scalar-matter coupling,
the quantity $\partial\rho_b/\partial\varphi\to
0$~\footnote{Recently, Sami and Tsujikawa analysed the model
numerically by considering $Q\to 0$~\cite{Tsujikawa:2006ph}. The
authors also studied the case $Q=$ const, but without modifying
the Friedmann constraint equation.} and hence
\begin{equation}\label{cons-eqn-minimal}
\Omega_b^\prime+2\epsilon \Omega_b+3(1+w_b)\Omega_b=0,
\end{equation}
where the prime denotes a derivative with respect to $\Delta N=\ln
[a(t)] +{\rm const}$ and $a(t)$ is the scale factor of a
four-dimensional Friedman-Robertson-Walker universe. In this case
the dynamics in a homogeneous and isotropic FRW spacetime may be
determined by specifying the field potential $V(\varphi)$ and/or
the scalar-GB coupling $f(\varphi)$.

For $A(\varphi)\ne {\rm const}$, there exists a new
parameter, $Q$; more precisely,
\begin{equation}
\frac{\partial\rho_m^{(i)}}{\partial\varphi}=- \eta \rho_m^{(i)}
Q_i, \quad \frac{\partial\rho_r}{\partial\varphi}=0.
\end{equation}
The variation in the energy densities of the ordinary matter,
$\rho_m^{(d)}$, and the relativistic fluid, $\rho_m^{(s)}$, and
their scalar couplings $Q_i$ are not essentially the same. Thus,
hereafter, we denote the $Q$ by $Q_d$ for an ordinary matter (or
dust) and $Q_s$ for a relativistic matter (or stiff fluid coupled
to $\varphi$). Equation (\ref{conservation-eq}) may be written as
\begin{eqnarray}
\Omega\Z{m, s}^\prime+2\epsilon \Omega\Z{m, s} + 3(1+w_m)
\Omega_{m, s} &=& - 3 \eta Q\Z{s}
\Omega_{m, s} \varphi^\prime,\label{Q-non-zero-1}\\
\Omega_{m, d}^\prime + 2\epsilon \Omega_{m, d} + 3 \Omega_{m, d}
&=& - 3 Q\Z{d}
\Omega\Z{m, d} \varphi^\prime,\label{Q-non-zero-2}\\
\Omega_r^\prime+ 2\epsilon \Omega_r+4 \Omega_r &=&
0.\label{Q-non-zero-3}
\end{eqnarray}
For a slowly rolling scalar (or dark energy) field $\varphi$,
satisfying $\Delta \varphi\propto \Delta N$ (where $N\equiv
\ln[a(t)]$ is the e-folding time), $Q_d$ is dynamically
determined, for any $f(\varphi)$, by the relation
\begin{equation}\label{non-mini-exact}
\Omega\Z{r}=Q_d \varphi^\prime {\Omega\Z{m, d}}.
\end{equation}
For $\Omega_r\sim \Omega\Z{m,d}$, the scalar field presumably
couples strongly to matter. However, for the present value of
$\Omega\Z{r} \sim 10^{-4}$ (on large cosmological scales), one
would hardly find a noticeable difference between the cases of
non-minimal and minimal coupling of $\varphi$ with matter.

In the following we adopt the convention $\kappa^2=8\pi G=1$,
unless shown explicitly.

\medskip
\subsection{Cosmological and astrophysical constraints}

It is generally valid that scalar-tensor gravity models of the
type considered in this paper are constrained by various
cosmological and astrophysical observations, including the big
bang nucleosynthesis bound on the $\varphi$-component of the total
energy density and the local gravity experiments. However, the
constraints obtained in a standard cosmological setup (for
instance, by analyzing the CMB data), which assumes general
relativity, that is, $Q=0$, cannot be straightforwardly applied to
the present model. The distance of the last scattering surface can
be (slightly) modified if the universe at the stage of BBN
contained an appreciable amount in energy in $\varphi$-component,
like $\Omega_\varphi\lesssim 1/6$ (see, e.g. \cite{Coc:2006}).

The couplings $Q_d$ and $Q_s$ can be constrained by taking into
account cosmological and solar system experiments. We should also
note that observations are made of objects that can be classified
as visible matter or dust~\cite{Damour:2002a,Damour:1990a}; hence
a value of $ Q_d^2 \ll 1$ is suggestive to agree with the current
observational limits on deviations from the equivalence principle.
If we also require $Q_s^2 \ll 1$, then the non-minimal coupling of
$\varphi$ with a relativistic fluid may be completely neglected at
present, since $\Omega_s\ll \Omega_d$. In this subsection, as we
are more interested on astrophysical and cosmological constraints
of the model, we take $\Omega_s=0$, without loss of generality.

Under PPN approximations~\cite{Martin:2005},
the local gravity constraints on $Q_d$ and
its derivative loosely imply that
\begin{equation}
m_{Pl}^2 Q_d^2 \lesssim 10^{-5}, \qquad
m_{Pl}\,|dQ_d/d\varphi|\lesssim {\cal O} (1).
\end{equation}
If $A(\varphi)$ is sufficiently flat near the current value of
$\varphi=\varphi_0$, then these couplings have modest effects on
large cosmological scales. Especially, in the case that
$A(\varphi) \propto e^{\zeta \kappa \varphi}$, the above
constraints may be satisfied only for a small $\zeta$ ($\ll 1$)
since $Q_d=\zeta$ is constant. On the other hand, if $A(\varphi)
\propto \cosh[\zeta\kappa\varphi]$, then $Q_d$ is only
approximately constant, at late times. Another possibility is that
$Q_d\propto e^{-\,\zeta\kappa \varphi}$; in this case, however,
for the consistency of the model, one also requires a steep
potential. In the particular case of $A(\varphi)\propto
e^{\zeta\kappa\varphi}$ with $\zeta\sim {\cal O}(1)$ (as one may
expect from string theory or particle physics beyond the standard
model), it may be difficult to satisfy the local gravity
constraints (under PPN approximation), unless there is a mechanism
like the one discussed in~\cite{Khoury:03aq} (see
also~\cite{Mota:2006fz}).

Only in the case $ Q_d \varphi^\prime \gtrsim 1$, is the last term
in (\ref{Q-non-zero-2}) possibly relevant, leading to a measurable
effect including a `fifth force causing violations of the
equivalence principle (see, e.g.,~\cite{Damour:2002a}). However,
in our model, it is important to realize that the coupling $Q_d$
is dynamically determined by the ratio
$\Omega_r/(\varphi^\prime\Omega_m)$. Even if we allow a small
slope for the time variation of the field, $\varphi^\prime\simeq
0.01$, which may be required for satisfying the BBN bound imposed
on the time variation of Newton's constant (see below), we find
$Q_d \sim 10^{-2}$ in the present universe, since $\Omega_r\sim
10^{-4}$ and $\Omega_m\sim 0.3$. In our numerical studies below we
shall allow $Q_d$ in a large range $10^{-5}\le Q_d^2 \lesssim
0.01$ and study the effect of $Q_d$ on the duration of matter (and
radiation) dominance (cf figure \ref{qv}).

One may be interested to give an order of magnitude to the scalar
field mass, $m_\varphi$. This is especially because the lighter
scalar fields appear to be more tightly constrained by local
gravity experiments as they presumably couple strongly to matter
fields. However, one of the important features of our model is
that the dynamics of the field $\varphi$ is not governed by the
field potential $V(\varphi)$ alone but rather by the effective
potential (cf equation (\ref{GB3})):
$$ V_{\rm eff}=V(\varphi)+\frac{1}{8} f(\varphi) {\cal R}^2 -
\eta \tilde{\rho}\Z{m} \int Q A(\varphi) d\varphi. $$ Here
$\tilde{\rho}\Z{m}\propto 1/a^3$, which is conserved with respect
to the Einstein frame metric $g_{\mu\nu}$. One can take $Q<0$, so
that the last term above contributes positively to the effective
potential; for the parametrization $A(\varphi)\propto
e^{\zeta\varphi/m_P}$, this requires $\zeta<0$. An important
implication of the above potential is that the mass of the field
$\varphi$, which is given by $m_\varphi$ ($\equiv \sqrt{V_{{\rm
eff},~\varphi\varphi}}$), it not constant, rather it depends on
the ambient matter density, $\tilde{\rho}_m$. It may well be that,
on large cosmological scales, $\varphi$ is sufficiently light,
i.e. $m_{\phi}\gtrsim 10^{-33} {\rm eV} \approx (10^{28} {\rm
cm})^{-1}$, and its energy density evolves slowly over
cosmological time-scales, such that $V_{\rm eff}\sim
V(\phi)+\frac{1}{8}\, f(\phi) {\cal R}^2 \sim {10}^{-47} ({\rm
GeV})^4$. However, in a gravitationally bound system, e.g. on
Earth, where the $\rho_m$ is some $10^{30}$ times larger than on
cosmological scales, that is $\rho_m \sim \rho_c \times 10^{30}
\sim 2~ {\rm g/cm^3}$, the Compton wavelength of the field may be
sufficiently small, $\lambda_c\sim m_\varphi^{-1} \lesssim 10^{3}
\, {\rm eV}^{-1} \sim 0.2~{\rm mm} $  as to satisfy all major
tests of gravity (see, e.g.~\cite{Khoury:03aq}, for similar
discussions).

There may also exist a constraint on the time variation of
Newton's constant. With the approximation that the matter is well
described by a pressureless (non-relativistic) perfect fluid with
density $\rho_m$, satisfying $\rho_r\ll \rho_m$, which implies
that $Q_d\ll  1$ or $A\simeq {\rm const}$ (cf
equation~\ref{non-mini-exact}), the growth of matter fluctuation
may be expressed in a standard form:
\begin{equation}
\ddot{\delta}+2\dot{\delta} H=4\pi G_{\rm eff} \, \rho_m \,\delta,
\end{equation}
where $\delta\equiv \delta\rho_m/\rho_m$. The effective Newton's
constant ${G}_{\rm eff} $ is given by~\cite{Amendola:2005}
\begin{equation}
{G}_{\rm eff} ={G} \left[1+3 \,\Omega_{GB}
-\frac{\dot{\varphi}}{H}\left(
\frac{\ddot{\varphi}}{\dot{\varphi}^2}+
\frac{f_{\varphi\varphi}}{f_\varphi}\right)\Omega_{GB} \right],
\end{equation}
where $\Omega_{GB}\equiv  H \dot{f} < 1$. In the particular case
of $f(\varphi)\propto e^{\alpha\varphi}$, this yields
\begin{equation}
{G}_{\rm eff} ={G} \left[1+ f
(2H\dot{\varphi}-\ddot{\varphi})\right].
\end{equation}
It is not improbable that ${G}_{\rm eff} \simeq G$ for the present
value of the field $\varphi_0$ and the coupling $f(\varphi_0)$,
since $\ddot{\varphi}/\dot{\varphi}\ll 1$ and $H \dot{f}<1$.
Indeed, $\dot{G}_{\rm eff}$ is suppressed (as compared to $G_{\rm
eff}$) by a factor of $\dot{\varphi}$ and thus one may satisfy the
current bound $\frac{d{G}_{\rm eff}}{dt}/{G}_{\rm eff} <
0.01\,H_0$, where $H_0$ is the present value of $H$, by taking
$\dot{\varphi}/H < 0.01$. This translates to the constraint
$|G_{\rm now}-G_{\rm nucleo}|/G_{\rm now}(t_{\rm now}-t_{\rm
nucleo}) < 10^{-12} {\rm yr}^{-1}$. Another opportunity for our
model to overcome constraints coming from GR is to have a coupling
$f(\varphi_0)$ which is nearly at its minimum.

The constraint on the time variation of Newton's constant may
arise even if $\Omega_{GB}=0$, given that $Q_d$ is nonzero. In
this case we find
$${G}_{\rm eff} = G A^2(\varphi) \left(1+ Q_d^2\right).$$
Nevertheless, in the present universe, since $\rho_r \ll \rho_m$,
$A(\varphi)\approx {\rm const}$ and $Q_d^2$ is small ($\ll 1$).

\section{Construction of inflationary potentials}

In this section we study the model in the absence of background
matter (and radiation). Equations (\ref{GB1})-(\ref{GB3}) form a
system of two independent equations of motions, as given by
\begin{eqnarray}
\gamma x^2 + 2y+ 6 \mu -6 &=& 0,\\
\mu^\prime+(\epsilon-1) \mu -\gamma x^2 - 2\epsilon &=&0.
\end{eqnarray}
The solution $\epsilon=0$ corresponds to a cosmological constant
term, for which $w=-1$~\footnote{The equation of state (EoS)
parameter $w\equiv \frac{p}{\rho}=-\frac{3 H^2+2\dot{H}}{3H^2}=
-1-\frac{2}{3}\,\epsilon$. The universe accelerates for $w<-1/3$,
or equivalently, for $\epsilon>-1$. In fact, for the
action~(\ref{gravi-eq}), it is possible to attain $w<-1$ or
equivalently $\epsilon>0$, without introducing a wrong sign to the
kinetic term.}. We shall assume that the scalar potential is
non-negative, so $y\ge 0$. Evidently, with $x^2>0$, as is the case
for a canonical $\varphi$, the inequality $\mu < 1$ holds at all
times. Here $\mu=1$ is a saddle point for any value of $\epsilon$.
Thus, an apparent presence of ghost states (or short-scale
instabilities) at a semi-classical level, as discussed
in~\cite{Calcagni:06ye}, which further extends the discussion
in~\cite{Hwang&Noh97}, is not physical. We shall return to this
point in more detail, in our latter discussions on small scale
instabilities incorporating superluminal propagations or ghost
states.

An interesting question to ask is: can one use the modified
Gauss-Bonnet theory for explaining inflation in the distant past?
In search of an answer to this novel question, it is sufficiently
illustrative
 to focus first on one simple example, by making the ansatz
\begin{equation}\label{two-ansatzs}
x \equiv  x_0\, e^{\alpha N/2}, \quad \mu \equiv \mu_0\, e^{\beta
N},
\end{equation}
where $x_0$, $\mu_0$, $\alpha$ and $\beta$ are all arbitrary
constants. The implicit assumption is that both $x$ and $\mu$
decrease exponentially with $N$, or the expansion of the universe.
We then find
\begin{equation}
\epsilon=\frac{2(\beta-1)\mu - \gamma x^2}{2(2-\mu)}.
\end{equation}
From this we can see that a transition between the $\epsilon>0$
and $\epsilon<0$ phases is possible if $\mu\ne 0$. Moreover, the
first assumption in (\ref{two-ansatzs}) implies that $\alpha N=
2\ln\varphi + \ln\varphi_1$,
where $\varphi_1$ is an arbitrary constant. 
The scalar potential is then given by
\begin{equation}\label{quartic}
V(\varphi)= H(\varphi)^2\,\left(3-\lambda_0\varphi^2-
\lambda_1\,\varphi^{\,2\beta/\alpha}\right),
\end{equation}
where $\lambda_0, \lambda_1$ are (arbitrary) constants. For a
canonical $\varphi$, so $x^2
>0$, we have $\lambda_0>0$, whereas the sign of $\lambda_1$ is determined
by the sign of $\mu_0$. Note that $\alpha=\beta$ is a special case
for which the potential takes the form $V(\varphi)\equiv
H^2(\varphi) \left(3- (\lambda_0+\lambda_1)\varphi^2\right)$. The
quantity $\epsilon$ (and hence $1+w$) cannot change its sign in
this case. Typically, if $\beta=2\alpha$, then the scalar
potential would involve a term which is fourth power in $\varphi$,
i.e., $V\propto H^2(\varphi) \,\varphi^4$. At this point we also
note that the potential~(\ref{quartic}) is different from a
symmetry breaking type potential $V\propto (\Lambda \pm
m_\varphi^2 \varphi^2 + \lambda_1 \varphi^4 + \cdots)$ considered
in the literature; here it is multiplied by $H^2(\varphi)$. An
inflationary potential of the form $V\propto
 \varphi^4$ is already ruled out by recent WMAP results, at
3$\sigma$-level, see e.g.~\cite{Peiris06ug}. In the view of this
result, rather than the monomial potentials, namely
$V(\varphi)\propto (\varphi/\varphi_0)^p$, a scalar potential of
the form $V(\varphi)\propto H^2(\varphi) (\varphi/\varphi_0)^p$,
with $p\ge 2$, as implied by the symmetries of Einstein's
equations, is worth studying in the context of inflationary
paradigm. Further details will appear elsewhere.

Another interesting question to ask is: is it possible to use the
modified Gauss-Bonnet theory to explain the ongoing accelerated
expansion of the universe? Before answering this question, we note
that, especially, at late times, the rolling of $\varphi$ can be
modest. In turn, it is reasonable to suppose that $\dot{\varphi}/H
\simeq $ const, or $x\simeq x_0$. Hence
\begin{equation}\label{potential-late time}
V(\varphi) =\frac{H^2}{2} \left( 6- \gamma {x_0}^2 + 6\mu_0
\,\e^{\beta\phi}\right)\equiv H(\varphi)^2 \left(\Lambda_0+
\Lambda_1\,\e^{\beta\phi}\right),
\end{equation}
where $\phi\equiv (\varphi-\varphi_0)/x_0$ and the Hubble
parameter is given by
\begin{equation}
H  =
H_0\,\left(1-\mu_0\,e^{\beta\phi}\right)^{(2+x_0)/2\beta-1}\,e^{-\,x_0\phi/2},
\end{equation}
with $H_0$ being an integration constant. Interestingly, a
non-vanishing $f(\varphi)$ not only supports a quartic term in the
potential, propotional to $H^2(\varphi)\,\varphi^4$, but its
presence in the effective action also allows the possibility that
the equation of state parameter $w$ switches its value between the
$w
> -1$ and $w < -1$ phases. We shall
analyse the model with the choice (\ref{potential-late time}) and
in the presence of matter fields, where we will observe that the
universe can smoothly pass from a stage of matter dominance to
dark energy dominance.


In the case where $\varphi$ is rolling with a constant velocity,
$\varphi^\prime\equiv c$, satisfying the power-law expansion
$a(t)\propto t^{1/p}$, or equivalently $H\equiv H_0 \,e^{-\,p N}$
and $p\ne 1$, we find that the scalar potential and the scalar-GB
coupling evolve as
\begin{eqnarray} V &=& V_0\, e^{-\, 2 p\phi}
+3(3p+1) f_1 H_0^2\, e^{\,(1- p) \phi},\\
f &=& f_0\,e^{\,2 p \phi} - \frac{f_1}{H_0^2}\, e^{\,(1+3p) \phi}+
f_2,
\end{eqnarray}
where $\phi\equiv (\varphi-\varphi_0)/{c}$, $f_1$ and $f_2$ are
arbitrary constants, and
\begin{equation}
V_0\equiv \frac{(6-6p+5c^2-pc^2)H_0^2}{2(1+p)},  \quad f_0\equiv
\frac{2p-c^2}{2p(p+1)H_0^2}.
\end{equation}
The potential is a sum of two exponential terms; such a potential
may arise, for example, from a time-dependent compactification of
$10$ or $11d$ supergravities on factor
spaces~\cite{Neupane:2005nb}. In general, both $V(\varphi)$ and
$f(\varphi)$ pick up additional terms in the presence of matter
fields, but they may retain similar structures. In fact, various
special or critical solutions discussed in the literature, for
example~\cite{Nojiri05b}, correspond to the choice $f_1=f_2=0$.

We can be more specific here. Let us consider the following
ansatz~\cite{Nojiri05b}:
\begin{equation}\label{asym-sol1}
a \propto (t+t_1)^\alpha, \qquad \varphi = \varphi_0 + \beta \ln
(t+t_1),
\end{equation}
for which obviously both $\epsilon$ and $\varphi^\prime$ are
constants:
\begin{equation} \epsilon \equiv \dot{H}/{H}^2=
-1/\alpha, \quad  \varphi^\prime \equiv \frac{\dot\varphi}{H}=
\frac{\beta}{\alpha}.
\end{equation}
For $\epsilon<0$, one can take $t_1=0$; the Hubble parameter is
given by
\begin{equation}
H=| \alpha| \,\e^{- (1/\beta)(\varphi-\varphi_0)}.
\end{equation} The scalar potential is
double exponential, which is given by
\begin{equation}
V= V_0\, \e^{-\,2\phi} + V_1\,\e^{(\alpha-1)\phi},
\end{equation}
where $\phi\equiv (\varphi-\varphi_0)/\beta$ and
\begin{equation}
 V_0\equiv
\frac{6\alpha^2(\alpha-1)+\beta^2(5\alpha-1)}{2(\alpha+1)}, \quad
V_1\equiv - 3 (\alpha+3)\alpha c_1.
\end{equation}
The scalar-GB coupling $f(\varphi)$ may be given by
\begin{equation}
 f =
f_0\,\e^{2 \phi}+ f_1\, \e^{(\alpha +3 )\phi} + f_2,
\end{equation}
with
\begin{equation}
f_0\equiv \frac{2\alpha-\beta^2}{2(1+\alpha)\alpha^2}, \quad f_1
\equiv \frac{c_1}{\alpha^2(\alpha+3)}, \quad f_2\equiv
\frac{c_2}{\alpha^2}.
\end{equation}
Of course, the numerical coefficient $f_2$ does not contribute to
Einstein's equations in four dimensions but, if non-zero, it will
generate a non-trivial term for the potential.

In the case $\epsilon >0$ (or $\dot{H}>0$), the above ansatz may
be modified as
\begin{equation}
a\propto (t_\infty-t)^\alpha, \qquad \varphi=\varphi_0 +\beta \ln
(t_\infty-t),
\end{equation}
where $\alpha<0$. The Hubble parameter is then given by $ H=
-\,\alpha(t_\infty-t)^{-1}$. Such a solution to dark energy is
problematic. To see the problem, first note that although this
solution avoids the initial singularity at $t=0$, it nonetheless
develops a big-rip type singularity at an asymptotic future
$t=t_\infty$. This is not a physically appealing case. Also the
above critical solution may be unstable under (inhomogeneous)
cosmological perturbations, which often leads to a super-luminal
expansion and also violates all energy conditions.

The reconstruction scheme presented, for example, in
ref.~\cite{Nojiri:2006je} was partly based on some special ansatz,
e.g. (\ref{asym-sol1}), which may therefore suffer from one or
more future singularities. However, as we show below, for the
model under consideration there exists a more general class of
cosmological solutions without any (cosmological) singularity.

\section{Inflationary cosmology: scalar-field dynamics}

Inflation is now a well established paradigm of a consistent
cosmology, which is strongly supported by recent WMAP
data~\footnote{The small non-uniformities observed for primordial
density or temperature fluctuations in the cosmic microwave
background provide support for the concept of inflation.}. It is
also generally believed that the small density (or scalar)
fluctuations developed during inflation naturally lead to the
formation of galaxies, stars and all the other structure in the
present universe. It is therefore interesting to consider the
possibility of achieving observationally supported cosmological
perturbations in low-energy string effective actions. For the
model under consideration, this can be done by using the standard
method of studying the tensor, vector, and scalar modes. We omit
the details of our calculations because they are essentially
contained in \cite{Hwang&Noh97} (see also
~\cite{Calcagni:06ye,Guo:2006ct}).

\subsection{Inflationary parameters}

One may define the slow roll parameters, such as $\epsilon_H$ and
$\eta_H$, associated with cosmological perturbations in a FRW
background, using two apparently different versions of the slow
roll expansion. The first (and more widely used) scheme in the
literature takes the potential as the fundamental quantity, while
the second scheme takes the Hubble parameter as the fundamental
quantity. A real advantage of the second
approach~\cite{Stewart:1993a} is that it also applies to models
where inflation results from the term(s) other than the scalar
field potential. Let us define these variables in terms of the
Hubble parameters~\footnote{The slow roll variable is
$\epsilon_H$, not $\epsilon$, the latter is defined by
$\epsilon\equiv H^\prime/H$.}:
\begin{equation}
\epsilon_H  \equiv 2
\left(\frac{H_\varphi}{H}\right)^2=\frac{2\epsilon^2}{{\varphi^\prime}^2},
\quad  \eta_H \equiv  \frac{ 2 H_{\varphi\varphi}}{H}=
\frac{2}{{\varphi^\prime}^2} \left(\epsilon^\prime+\epsilon^2
-\epsilon \frac{\varphi^{\prime\prime}}{\varphi^\prime}\right)
\end{equation}
(in the units $\kappa=1$). Here, as before, the prime denotes a
derivative w.r. to e-folding time $N\equiv \ln[a(t)]$, not the
field $\varphi$. One also defines the parameter $\xi_H$, which is
second order in slow roll expansion:
\begin{equation}
\xi_H \equiv \frac{1}{2} \left(\frac{H_\varphi
H_{\varphi\varphi\varphi}} {H^2}\right)^{1/2}=\left(\epsilon_H
\eta_H -\sqrt{2\epsilon_H}\,
\frac{\eta_H^\prime}{\varphi^\prime}\right)^{1/2}.
\end{equation}
These definitions are applicable to both cases, $V(\varphi)\ne 0$
and $V(\varphi)=0$, and are based on the fact that inflation
occurs as long as $\frac{\dd}{\dd t} (\frac{1}{a H})< 0$ holds.
The above quantities ($\epsilon_H, \eta_H, \xi_H$) are known as,
respectively, the slope, curvature and jerk parameters.

Typically, in the case $f(\varphi)=0$, or simply when
$|\Omega_{GB}| \ll \Omega_\varphi$, so that the coupled GB term
becomes only subdominant to the field potential, the spectral
indices of scalar and tensor perturbations to the second order in
slow roll expansion may be given by~\cite{Stewart:1993a} (see
also~\cite{Peiris06ug})
\begin{eqnarray}
n_{\cal R}-1 &=& -4 \epsilon_H + 2
\eta_H-2(1+c)\epsilon_H^2-\frac{1}{2}(3-5c)
\epsilon_H\eta_H+\frac{1}{2}(3-c)\xi_H^2,\nonumber\\
n_T &=& -2\epsilon_H - (3+c)\epsilon_H^2+(1+c)\epsilon_H\eta_H,
\end{eqnarray}
where $c=4(\ln 2+\gamma)-5 \approx 0.08$. For the solutions
satisfying $\xi_H\simeq 0$, implying that both $\epsilon_H$ and
$\eta_H$ are much smaller than unity (at least, near the end of
inflation) and their time derivatives are negligible, we have
\begin{equation}\label{indices-Hubble}
n_{\cal R}-1\simeq  -4\epsilon_H + 2\eta_H, \qquad n_T\simeq -2
\epsilon_H.
\end{equation}
In fact, $n_{\cal R}$ and $n_T$, along with the scalar-tensor
ratio $r$, which is given by $r\approx
16\epsilon_H+32c(\epsilon_H-\eta_H)\epsilon_H$, are the quantities
directly linked to inflationary cosmology.

Below we write down the results directly applicable to the
theories of scalar-tensor gravity with the Gauss-Bonnet term. Let
us introduce the following quantities:~\footnote{The parameter
$\epsilon_3$ defined in~\cite{Hwang&Noh97} (see
also~\cite{Guo:2006ct}) is zero in our case since the
action~(\ref{effective}) is already written in the Einstein frame,
and $\varphi$ does not couple with the Ricci-scalar term in this
frame.}
\begin{eqnarray}
&&\epsilon_1 =-\frac{\dot{H}}{H^2}= -\epsilon, \qquad \epsilon_2 =
\frac{\ddot{\varphi}}{\dot{\varphi}H}=\frac{x^\prime+\epsilon
x}{x}, \qquad \epsilon_4 =
\frac{\varrho^\prime}{2\varrho},\nonumber
\\
&& \epsilon_5 \equiv \frac{\mu}{2(1-\mu)}, \qquad \epsilon_6
\equiv -\frac{\mu^\prime}{2(1-\mu)}.
\end{eqnarray}
where
\begin{equation}
\varrho \equiv \gamma+\frac{3\mu^2}{2(1-\mu)x^2}, \qquad \mu\equiv
\dot{f} H=f^\prime H^2.
\end{equation}
Even in the absence of the GB coupling ($\mu=0$), hence
$\epsilon_4=\epsilon_5=\epsilon_6=0$, there are particular
difficulties in evaluating the spectral indices $n_{\cal R}$ and
$n_T$, and the tensor-scalar ratio $r$, in full generality. In
perturbation theory it is possible to get analytic results only by
making one or more simplifying assumptions. In the simplest case,
one treats the parameters $\epsilon_i$ almost as constants, so
their time derivatives are (negligibly) small as compared to other
terms in the slow roll expansion. An ideal situation like this is
possible if inflation occurred entirely due to the power-law
expansion, $a(\tau)\propto |\tau|^{-\,1/(1+\epsilon)}$, where the
conformal time $\tau\equiv -1/[aH(1+\epsilon)]$. In this case, the
spectral indices of scalar and tensor perturbations are well
approximated by
\begin{equation}
n_{\cal R}-1=3-\Big\arrowvert
\frac{3+\epsilon_1+2\epsilon_2+2\epsilon_4}{1-\epsilon_1}\Big\arrowvert,
\quad n_T=3- \Big\arrowvert
\frac{3-\epsilon_1+2\epsilon_6}{1-\epsilon_1}\Big\arrowvert.
\label{spectral-indices}
\end{equation}
Note that not all $\epsilon_i$ are smaller than unity.
Nevertheless, an interesting observation is that, for various
explicit solutions found in this paper, the quantity $\epsilon_4$
is close to zero, while $\epsilon_5$, $\epsilon_6$ can have small
variations during the early phase of inflation. But, after a few
e-folds of inflation, $\Delta N\gtrsim 5$, these all become much
smaller than unity, so only the first two terms ($\epsilon_1$,
$\epsilon_2$) enter into any expressions of interest. In any case,
below we will apply the formulae (\ref{spectral-indices}) to some
explicit cosmological solutions.

The above relations are trustworthy only in the limit where the
speeds of propagation for scalar and tensor modes, which may be
given by
\begin{equation}
c_{\cal
R}^2=1+\frac{\mu^2[4\epsilon(1-\mu)+\ddot{f}-\mu]}{(1-\mu)[2\gamma(1-\mu)
x^2+3\mu^2]},\quad c_{T}^2=\frac{1-\ddot{f}}{1-\dot{f}
H}=\frac{1-\mu^\prime +\epsilon \mu}{1-\mu},
\end{equation}
where $\mu=\dot{f} H$ and $x=\dot{\varphi}/H$, take approximately
constant values. These formulae may be expressed in terms of the
quantity $\nu$, which is defined below in (\ref{def-for-nu}),
using the relation $\mu=\nu^\prime-2\epsilon\nu$. The propagation
speeds depend on the scalar potential only implicitly, i.e.,
through the background solutions which may be different for the
$V(\varphi)=0$ and $V(\varphi)\ne 0$ cases. In the case where
$c_{\cal R}^2$ and $c_T^2$ are varying considerably, the
derivative terms like $\dot{\epsilon_5}$, $\ddot{\epsilon_5}$ are
non-negligible, for which there would be non-trivial corrections
to the above formulae for $n_{\cal R}$ and $n_T$. Furthermore, the
spectral indices diverge for $\epsilon_1\sim 1$, thus the results
would apply only to the case where $\epsilon_1\ll 1$. In this
rather special case, which may hold after a few e-folds of
power-law expansion through to near the end of inflation, we find
that the tensor-to-scalar ratio is approximately given by
\begin{equation}
r\equiv \frac{A_T^2}{A_{\cal R}^2}\approx 16 \frac{2\gamma
x^2(1-\mu)+3\mu^2}{(2-\mu)^2} \left(\frac{c_{\cal
R}}{c_T}\right)^{3}.
\end{equation}
This is also the quantity directly linked to observations, other
than the spectral indices $n_{\cal R}$ and $n_T$. The WMAP data
put the constraint $r<0.55$ at $2\sigma$ level.

\subsection{Inflating without the potential, $V(\varphi)=0$}

In this subsection we show that in our model it is possible to
obtain an inflationary solution even in the absence of the scalar
potential. It is convenient to introduce a new variable:
\begin{equation}\label{def-for-nu}
\nu(\varphi) \equiv f(\varphi) H^2.
\end{equation}
This quantity should normally decrease with the expansion of the
universe, so that all higher-order corrections to Einstein's
theory become only sub-leading~\footnote{Particularly in the case
$\nu\simeq $ const, the coupled Gauss-Bonnet term $\frac{1}{8}
f(\varphi) {\cal R}_{GB}^2 \equiv 3 f(\varphi) H^4 (1+\epsilon)= 3
\nu_0 H^2(1+\epsilon)$ may be varying being proportional to the
Einstein-Hilbert term, $R/2=3H^2(2+\epsilon)$.}. In the absence of
matter fields, and with $V(\varphi)=0$, the equations of motion
reduce to the form
\begin{eqnarray}\label{zero-V-2nd-order}
&& x^\prime=-\frac{6(1+\epsilon)+(5+\epsilon)\gamma x^2}{2\gamma x}, \nonumber \\
&& \nu^\prime= 1+ 2\epsilon \nu -\frac{\gamma}{6} x^2.
\end{eqnarray}
Fixing the (functional) form of $\nu(\varphi)$ would alone give a
desired evolution for $\epsilon$. Notice that for $x=$ const, one
has $\epsilon=$ const; such a solution cannot support a transition
between acceleration and deceleration phases. It is therefore
reasonable to suppose that $x\ne $ const, and also that $\nu$
varies with the number of e-folds, $N$. Take
\begin{equation}
\nu \propto \e^{ m N},
\end{equation}
where $m$ is arbitrary. As the simplest case, suppose that $m
\simeq 0$ or equivalently $\nu \simeq $ const $\equiv \nu_0$. In
this case we find an explicit exact solution which is given by
\begin{equation}\label{soln-epsilon}
\epsilon=-\beta_0 + \beta \tanh\left(\beta \Delta N\right),
\end{equation}
where $\Delta N= N+N_0$, with $N_0$ being an integration
constant~\footnote{One may choose the constant $N_0$ differently.
Here we assume that the scale factor before inflation is $a_0$, so
initially $N\equiv \ln (a(t)/a_0)=0$. As the universe expands,
$N\gg 0$, since $a(t)\gg a_0$. This last assumption may be
reversed, especially, when one wants to use the model for studying
a late time cosmology, where one normalizes $a(t)$ such that its
present value is $a_0$, which implies that $N<0$ in the past.},
and
\begin{equation}
\beta_0 \equiv \frac{5 \nu_0 +1}{2 \nu_0}, \quad \beta \equiv
\frac{1}{2\nu_0} \sqrt{25\nu_0^2-2\nu_0+1}
\end{equation}
Clearly, the case $\nu_0=0$ must be treated separately. The
evolution of $\varphi$ is given by $\varphi^\prime =\pm
\sqrt{(6+12 \epsilon\nu_0)/\gamma}$. The Hubble parameter is given
by
\begin{equation}
H=H_0\,\e^{-\beta_0 N} \cosh(\beta \Delta N).
\end{equation}
For $\nu_0 >0$, both $\beta_0$ and $\beta$ are positive and
$\beta_0 > \beta$, implying that Hubble rate would decrease with
the number of e-folds. In turn, the EOS parameter $w$ is greater
than $-1$. However, if $\nu_0<0$, then we get $\beta<0$, whereas
$\beta_0 <0$ for $-0.2<\nu_0 <0$ and $\beta_0
> 0$ for $\nu_0 <-0.2$. The $v_0<0$ solution
clearly supports a super-luminal expansion, see also
\cite{Kanno:2006ty}.

With the solution (\ref{soln-epsilon}), we can easily obtain
$50-60$ e-folds of inflation (as is required to solve the horizon
and the flatness problem) by choosing parameters such that
$1+\beta-\beta\Z0>0$. An apparent drawback of this solution is
that it lacks a mechanism for ending inflation. We will suggest
below a scenario where this problem may be overcome.

\begin{figure}[ht]
\begin{center}
\epsfig{figure=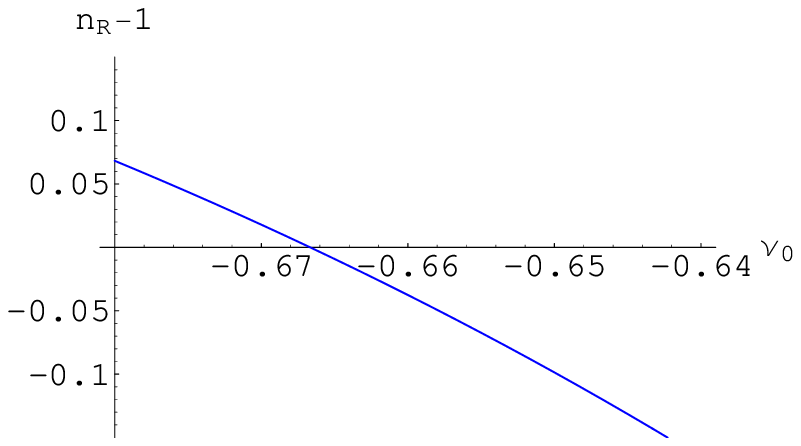,height=2.0in,width=2.8in}
\hskip0.2in
\epsfig{figure=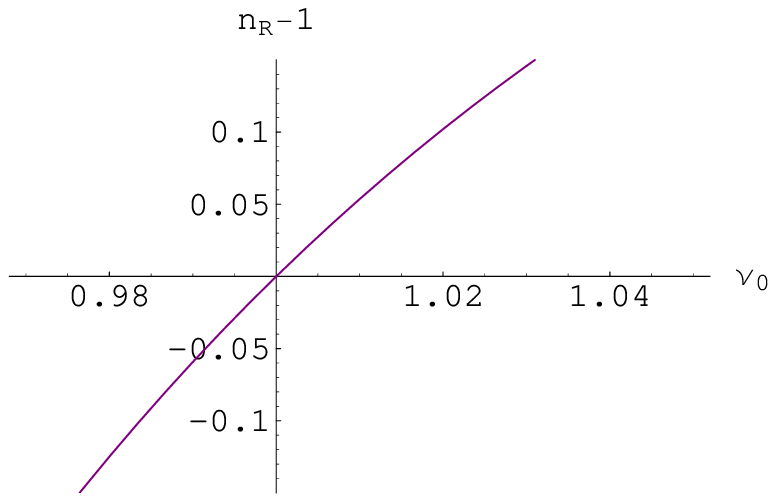,height=2.0in,width=2.8in}
\end{center}
\caption{The index $n_{\cal R}-1$ as the function of $\nu_0$, with
$V(\varphi)=0$ (cf equation 4.13); $n_{\cal R}$ does not vary much
with the number of e-folds except that it is blue-tilted for
$\Delta N \lesssim 0$. Even though $\nu_0<0$ generally implies a
super-luminal inflation, it is possible to achieve a red-tilted
spectrum for $\nu_0>-2/3$.} \label{figure1}
\end{figure}

The constraint on the spectral indices might give a stringent
bound on the scalar-Gauss-Bonnet coupling $f(\varphi)$, or the GB
density fraction, $\Omega_{GB}$. The scalar index satisfies the
bounds provided by the recent WMAP results, namely $n_{\cal
R}=0.958_{-0.018}^{+0.014}$, for both signs of $\nu$, but in a
limited range, see Fig.~\ref{figure1}. In the $V(\varphi)=0$ case,
in order to end inflation the function $\nu(\varphi)$ could be
modified in a suitable way; in the case $m N<0$, implying that
$\nu(\varphi)$ decreases with $N$, inflation ends naturally. With
this particular example, we do not wish to imply that the model
with a vanishing potential alone is compatible with recent data.
Such a model may suffer from one or more (quantum) instabilities
at short distance scales and/or local GR
constraints~\cite{Amendola:2005} once the model is coupled to
matter fields. As in ~\cite{Farese:2000a}, we find that a model
with $V(\varphi)>0$ is phenomenologically more viable than with
$V(\varphi)=0$.

As noted above, a super-luminal expansion may occur if $\nu_0<0$,
which corresponds to the case $\delta<0$ studied before
in~\cite{Antoniadis93a}. However, unlike claimed there, we find
that it is possible to get a cosmic inflation of an arbitrary
magnitude without violating the (null energy) condition
$\dot{H}/H^2\le 0$. For the solution (\ref{soln-epsilon}), this
simply required $\nu_0>0$; the Hubble rate would decrease with
$N$. For $\nu_0<0$, one gets $\dot{H}/H^2>0$, especially, when
${\Delta N}\gtrsim 0$. The precise value of redshift $z$ ($\equiv
e^{\Delta N}-1$) with such a drastic change in the background
evolution actually depends on the strength of the coupling
$f(\varphi)$, or the time variation of $\varphi$.

\subsection{Inflating with an exponential potential}

Let us take the scalar potential of the form
\begin{equation}\label{specific-potential}
V(\varphi)=V_0\,e^{- \,2\varphi/\varphi_0},
\end{equation}
leaving $f(\varphi)$ unspecified. The system of autonomous
equations is then described by
\begin{eqnarray}
&& \frac{d y}{dN} = -2 y \epsilon -\frac{2 x y}{\varphi_0},\label{expo-poten1}\\
&& \frac{d x}{dN}=\frac{(1+\epsilon)(y-3)}{\gamma x} -
\frac{2y}{\gamma\varphi_0}-\frac{(5+\epsilon) x}{2},\label{expo-poten2} \\
&& u= \frac{1}{x}-\frac{y}{3x}-\frac{\gamma
x}{6}.\label{expo-poten3}
\end{eqnarray}
These equations admit the following de Sitter (fixed point)
solution
\begin{equation}\label{de-Sitter-fixed}
x=0, \quad V= \Lambda_0, \quad H=\sqrt{\frac{\Lambda_0}{3}}, \quad
u=u(\varphi),
\end{equation}
which corresponds to the case of a cosmological constant term, for
which $w=-1$. In the particular case that $u\propto x$, so that $u
\equiv f_{,\,\varphi}H^2$ is varying (decreasing) being
proportional to the kinetic term for $\varphi$, we find an
explicit solution, which is given by
\begin{equation}
x =\varphi_0 -\alpha\,\varphi_0 \tanh\alpha (N+N\Z1), \quad u =
-\frac{\gamma x}{3}, \quad \epsilon=\frac{2\gamma x^2\varphi_0-2
x(6+\gamma x^2)}{\varphi_0(6+\gamma x^2)},\label{second-soln}
\end{equation}
where $\alpha \equiv
\sqrt{(\gamma\varphi_0^2-6)/\gamma\varphi_0^2}<1$ and $N\Z1$ is an
integration constant. This solution supports a transition from
deceleration ($\epsilon<-1$) to acceleration ($\epsilon>-1$), but
inflation has no exit. The model should be supplemented by an
additional mechanism, allowing the field $\varphi$ to exit from
inflation after $50 - 60$ e-folds of expansion (see a discussion
below).

For a negligibly small kinetic term, $x\simeq 0$, we have
$\Omega_{GB}\propto x^2 \approx 0$, while for a slowly rolling
$\varphi$, $\Omega_{GB} \ne 0$. This behaviour is seen also from
our numerical solutions in the next section. The evolution of
$\varphi$ is given by
\begin{equation}
\varphi = \int x\, dN = N\varphi_0 -\varphi_0\,\ln \cosh \alpha
(N+N\Z1)+{\rm const}.
\end{equation}
The scale factor, as given by $a \propto \exp[\int H\, dt] \propto
\exp\left[ \int \exp[\int \epsilon\,dN]\, dt\right]$, is regular
everywhere. The coupling $f(\varphi) \equiv \int
[u(\varphi)/\varphi^\prime H^2] \,d\varphi$ is rather a
complicated function of $\varphi$, which is generally a sum of
exponential terms. But given the fact that $x \simeq $ const at
late times, or when $|\alpha (N+N\Z1)|\gtrsim 2$, we get
$f_{,\,\varphi} \propto e^{\,2\varphi/\varphi_0}$. It is not
difficult to see that the special solution considered
in~\cite{Nojiri05b} corresponds to this particular case. To see it
concretely, take
\begin{equation}\label{specific-coupling}
f(\varphi)=f_0\,e^{2\varphi/\varphi_0},
\end{equation}
or, equivalently $\varphi_0 u^\prime-2u(\varphi_0\epsilon+x)=0$.
For the solution (\ref{second-soln}) this last condition imposes
the constraint $\gamma \varphi_0^2=6$ and hence
\begin{equation}\label{critical-sol}
x=x_0, \quad u=-\gamma x_0/3. \end{equation} The choice made
in~\cite{Nojiri05b} for $V(\varphi)$ and $f(\varphi)$ thus
over-constrains the system unnecessarily.

\begin{figure}[ht]
\begin{center}
\hskip-0.3cm
\epsfig{figure=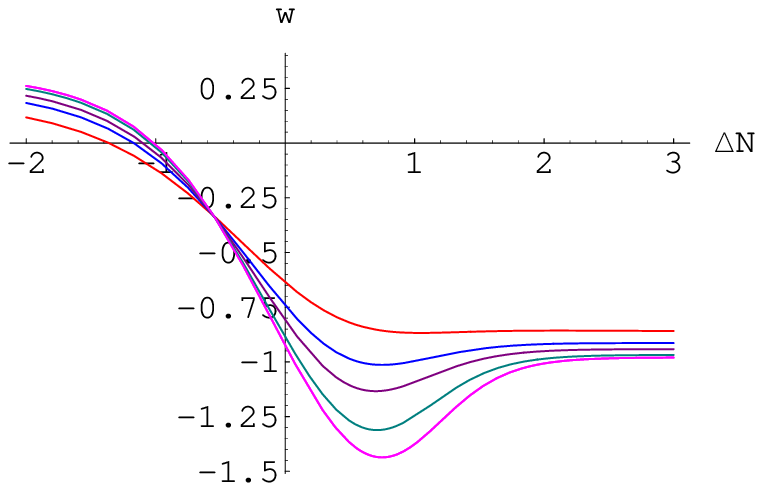,height=2.2in,width=3.0in}
\hskip0.2cm
\epsfig{figure=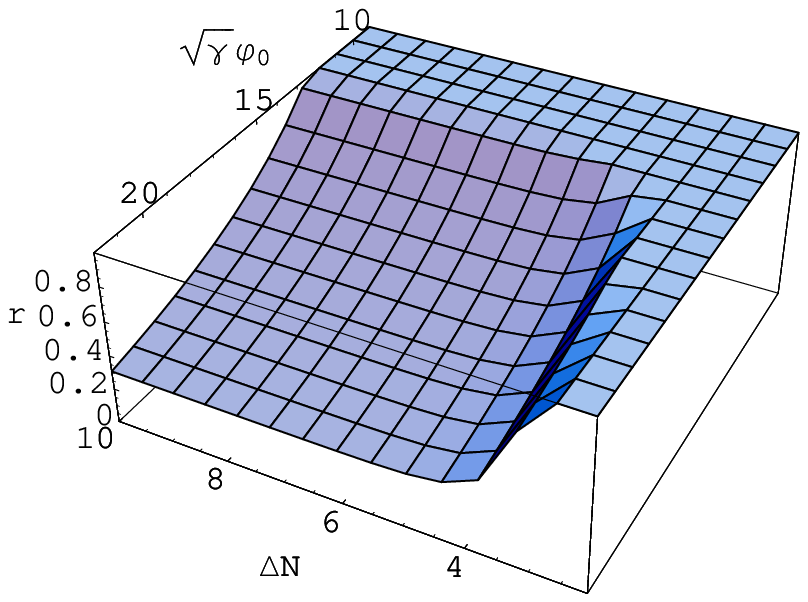,height=2.2in,width=3.0in}
\end{center}
\caption{The left panel shows that the equation of state parameter
$w$ that momentarily crosses the value $w=-1$. From top to bottom
$\sqrt{\gamma} \varphi_0=4, 5, 6, 8$ and $10$. The right panel
shows the tensor-to-scalar ratio of cosmological perturbations; to
adhere to the experimental constraint coming from recent WMAP
results, namely $r<0.5$, one might require an almost flat
potential, $\sqrt{\gamma} \varphi_0 \gtrsim 16$.}
\label{crossing1}
\end{figure}

\begin{figure}[ht]
\begin{center}
\epsfig{figure=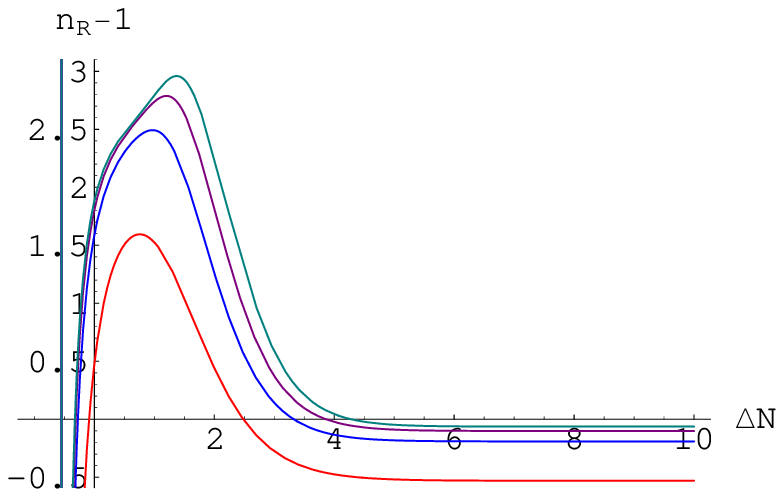,height=2.0in,width=2.8in}
\hskip0.15in
\epsfig{figure=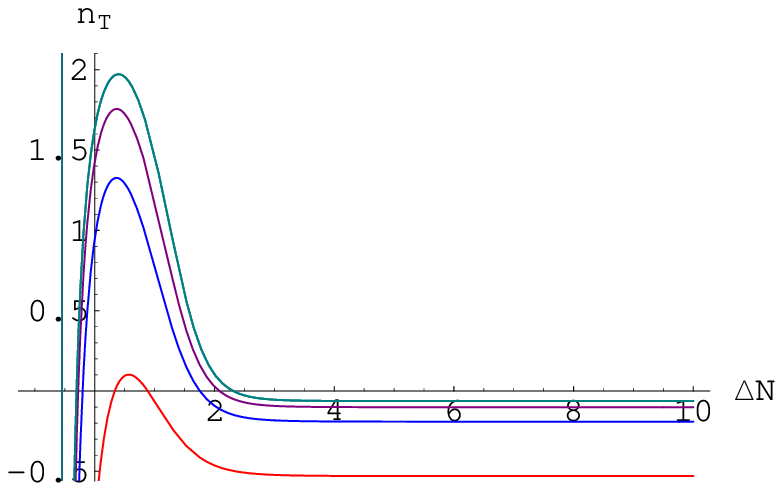,height=2.0in,width=2.8in}
\end{center}
\caption{The spectral indices $n_{\cal R}-1$ and $n_T$ evaluated
for the solution (4.21) by using the formulae given in (4.7); from
top to bottom $\sqrt{\gamma}\varphi_0=10, 8, 6$ and $4$.}
\label{spetral-indices}
\end{figure}

Can the solution~(\ref{second-soln}) be used for explaining early
inflation? In order to answer this question concretely, one may
wish to evaluate cosmological parameters such as $n_{\cal R}$ and
$n_T$. We can get the value of $n_{\cal R}$ in the range $0.94
\lesssim n_{\cal R} \lesssim 0.99$ by taking $9\lesssim
\sqrt{\gamma}\varphi_0\lesssim 20$. In fact, $n_{\cal R}$ may
exceed unity precisely when $\epsilon>0$, leading to a momentary
violation of the null energy condition (cf figure \ref{crossing1}
(left panel)). However, after a certain number of e-folds of
expansion, the tensor modes or the gravitational waves damp away
(cf figure \ref{crossing1} (right panel)), leading to the
observationally consistent result $|n_T|\lesssim 0$ and $n_{\cal
R}-1\lesssim 0$ - leading to slightly red-tilted spectra (cf
figure \ref{spetral-indices}).


For the validity of (\ref{spectral-indices}), we require not only
$\dot{\epsilon_i}\ll 1$ but also that the speed of propagation for
scalars and tensor modes, $c_k^2$ and $c_T^2$, are nearly
constant. This is however not the case for the above solution when
$0\lesssim \Delta N \lesssim 3$, for which both $c_r^2$ and
$c_T^2$ change rapidly. That means, the relations
(\ref{spectral-indices}) must be modified in the regime where
$c_{\cal R}^2, c_T^2$ are varying. After a few number of e-folds
of expansion, we have $c_{\cal R}^2, c_T^2 \simeq $ const. In this
limit the spectral indices are found to be slightly red-tilted,
i.e. $n_{\cal R}-1\lesssim 0$, for $\sqrt{\gamma}\varphi_0 \gtrsim
8$.

A few remarks may be relevant before we proceed. The WMAP
constraints on inflationary parameters that we referred to above,
such as, the scalar index $n_{\cal R}$ and the tensor-to-scalar
ratio $r$, are rather model dependent and may differ, for
instance, for a model where the ``inflaton" field is non-minimally
coupled to gravity, or the spacetime curvature tensor. Hence,
although those values may be taken as useful references, they may
not be strictly applicable to all inflationary solutions driven by
one or more scalar fields. In fact, some of our cases may have
more freedom (less tight constraints) than the canonical models
with a scalar potential proportional to $\varphi^4$ or
$m_\varphi^2\varphi^2$, and $f(\varphi)=0$. We leave for future
work the implications of WMAP data to scalar potentials of the
form $V\propto H^2(\varphi) \varphi^n $ ($n\ge 2$).

\subsection{Inflating with an exponential coupling}

\begin{figure}[ht]
\begin{center}
\epsfig{figure=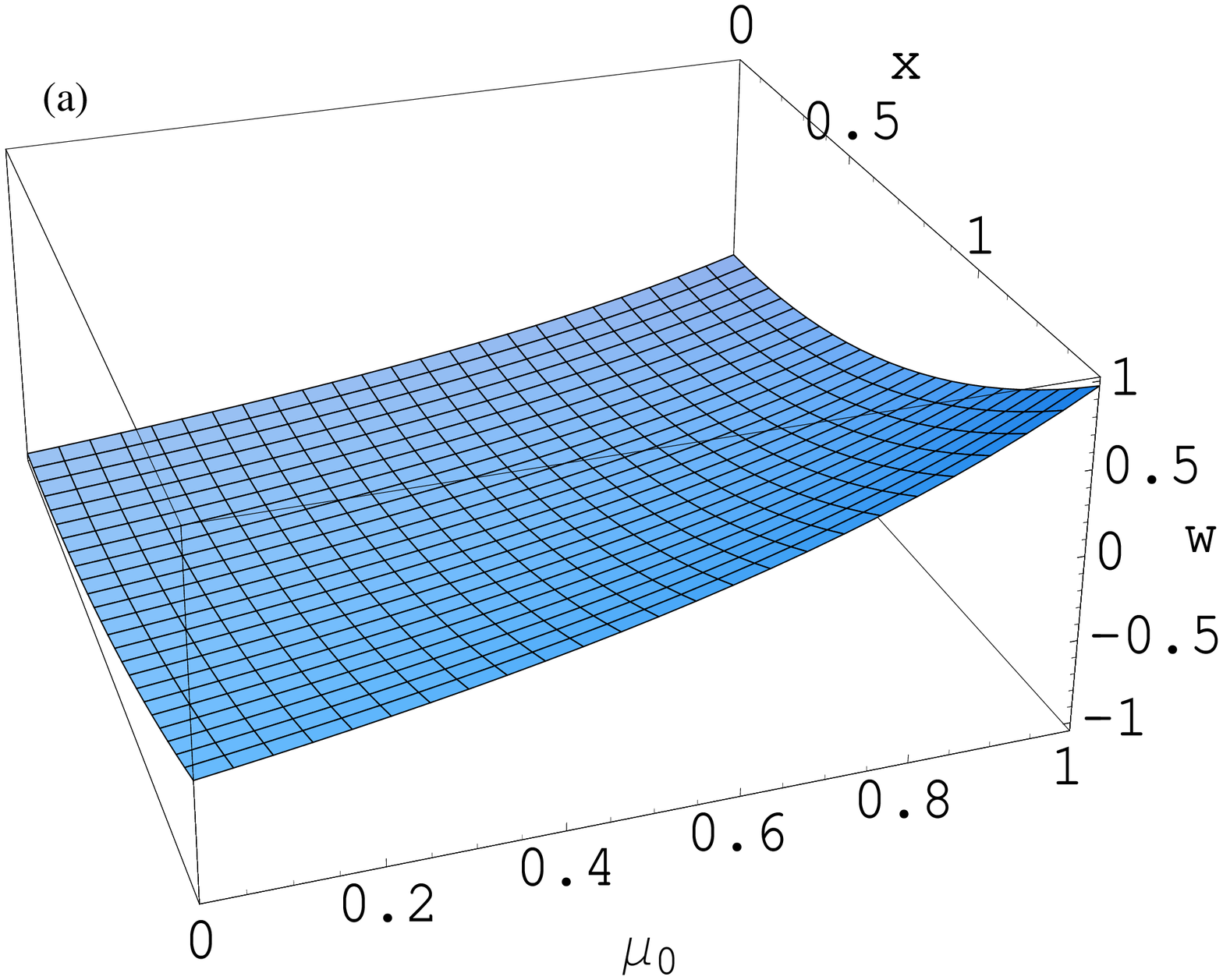,height=2.0in,width=2.8in}
\hskip0.2in
\epsfig{figure=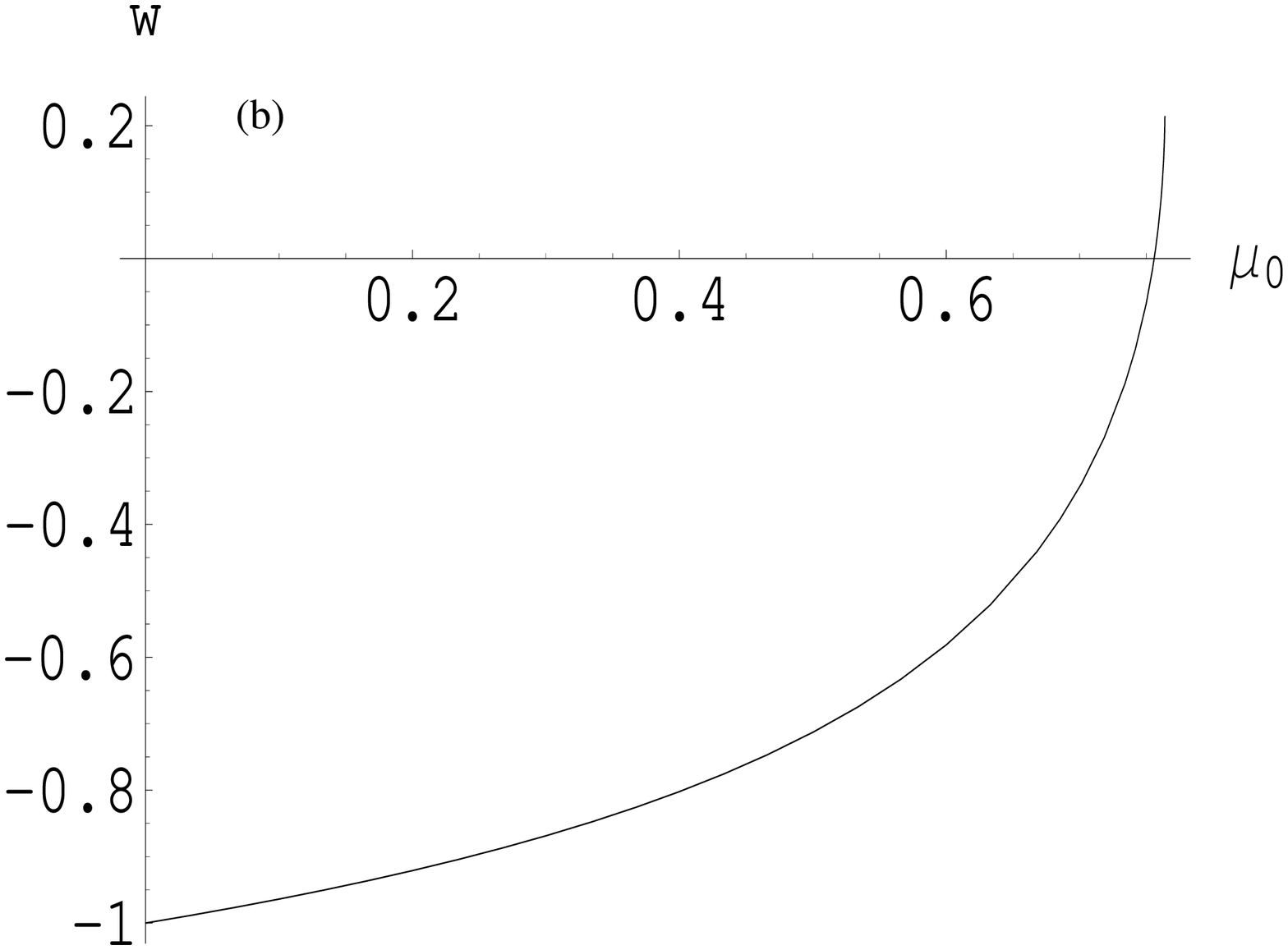,height=2.0in,width=2.8in}
\end{center}
\caption{The left panel (a) shows the initial equation of state
parameter ${\rm w}$ with various initial conditions for
$\Omega_{GB} = \mu_0$ and $x$, while, the right panel (b) shows a
possible variation of $w$ for a particular value of $\mu_0$;
solutions evolve to these values regardless of $x_{ini}$ provided
the fixed points exist and $0 < x_{\rm ini} < x_2$. We have taken
$\varphi_0^{-1} = 2$ and $\gamma=1$} \label{uxk}
\end{figure}

Let us take the scalar-GB coupling of the following form:
\begin{equation}\label{specific-coupling}
f_{,\varphi}=f_0\,e^{2\varphi/\varphi_0},
\end{equation}
but without specifying the potential, $V(\varphi)$. With this
choice, the system of monotonous equations is given by
\begin{eqnarray}
\frac{du}{dN}&=&\frac{2(x+\epsilon \varphi_0)}{\varphi_0}\,u,\\
 \frac{dx}{dN} &=&\frac{2\epsilon+\gamma x^2}{u} +(1-3\epsilon)
x-\frac{2x^2}{\varphi_0},\\
 y &= &3-\frac{\gamma}{2}\,x^2-3 u x.
\end{eqnarray}
The de Sitter fixed point solution for which $x=0$ is the same as
(\ref{de-Sitter-fixed}). In the following we consider two special
cases: \\
(1) Suppose that $\mu \equiv {\rm const}$, that is,
$\Omega_{GB} = $ const $\equiv \mu_0$. Then we find
\begin{equation}
\epsilon=\frac{\mu_0 +\gamma x^2}{\mu_0-2}, \quad
\frac{dx}{dN}=\frac{2(\gamma\varphi_0 x^3-(2-\mu_0)x^2+2\mu_0
\varphi_0 x)}{(2-\mu_0)\varphi_0}.
\end{equation}
These equations may be solved analytically for
$\mu_0=2/3$~\footnote{This is actually a critical point in the
phase space, which may be seen also in cosmological perturbation
analysis, see, e.g.,~\cite{Koivisto:2006}.}; as the solutions are
still messy to write, we only show the behavior of $w$ in
Mathematica plots. In the next section we will numerically solve
the field equations, in the presence of matter, allowing us to
consider all values of $\mu_0$. We should at least mention that
the above system of equations has a pair of fixed point solutions:
\begin{equation}
x_1= {2 - \mu_0 - \sqrt{(2-\mu_0)^2- 8 \gamma \varphi_0 \mu_0}
\over 2 \gamma \varphi_0}, \quad x_2 = {2 - \mu_0 +
\sqrt{(2-\mu_0)^2 - 8 \gamma \varphi_0 \mu_0} \over 2 \gamma
\varphi_0}.
\end{equation}
The fixed point $x_1$ is an attractor, while the $x_2$ is a
repeller provided that $x_1, x_2$ are real. In the case where $8
\gamma \varphi_0 \mu_0 > (2-\mu_0)^2$ the solution diverges. The
solution also diverges for initial values of $x$ such that $x < 0$
and $x > x_2$. If these conditions are not violated, the solutions
would always converge to the attractor fixed point $x=x_1$. The
evolution of the solution is monotonic from the initial value of
$\epsilon$ to the $\epsilon$ given by the attractor fixed point at
$x=x_1$.  The initial value of $w$ for a wide range of $\mu_0$ and
initial values of $x$ may be read from figure \ref{uxk}(a), and
evolve to the $w$ given in figure \ref{uxk}(b) for a specific
value of $\mu_0$.

(2) Next suppose that $u=$ const $\equiv u_0$, instead of $ux
=\mu$ =const. In this case, the quantity $x$ decreases quickly
with the expansion of the universe; the GB density fraction,
$\Omega_{GB}$, also decreases with the number of e-folds. The
explicit solution is
\begin{equation}
\epsilon=-\frac{x}{\varphi_0},\qquad
x=\frac{2-u_0\varphi_0}{\gamma\varphi_0+ u_0+(2-u_0\varphi_0)
u_1\,e^{(2/u_0\varphi_0-1)N}},
\end{equation}
where $u_1$ is an integration constant. Without any surprise, as
$x\to 0$, $\epsilon\to 0$; the model then corresponds to the
cosmological constant case.

Inflation is apparently future eternal for most solutions
discussed above, e.g.~(\ref{soln-epsilon}) and
(\ref{second-soln}). However, in a more viable cosmological
scenario, the contribution of matter field may not be completely
negligible. This is because, during inflation, the field $\phi$,
while slowly evolving down its potential, which satisfies
$\Delta\varphi\propto \Delta N$, decays into lighter particles and
radiation. The inflaton may even decay to heavier particles,
especially, if the reheating of the universe was due to an
instance preheating~\cite{Felder98}. In turn, a significant amount
of the energy density in the $\varphi$ component might be
transforming into the radiation and (baryonic plus dark) matter,
with several hundreds of degrees of freedom and with all
components present, e.g. stiff matter ($p=\rho$) and radiation
($3p=\rho$). In turn, the slow roll type parameter $\epsilon$
would receive a non-trivial contribution from the matter fields.
Explicitly, we find
\begin{equation}
\epsilon =- \frac{3}{2}\,(1+w) \Omega\Z{b}
-\frac{{\varphi^\prime}^2}{2},
\end{equation}
where $\Omega_b= \Omega\Z{m}+\Omega\Z{r}$ and $w\equiv
(p\Z{m}+p\Z{r})/(\rho\Z{m}+\rho\Z{r})$. Inflation ends when there
is significant portion of matter fields, which makes $\epsilon$
more and more negative and hence $\epsilon< -1$.

Before proceeding to next section we also wish to make a clear
separation between the inflationary solutions that we discussed
above and the dark energy cosmologies that we will discuss in the
following sections. Although it might be interesting to provide a
natural link between the early universe inflation and the late
time cosmic acceleration attributed to dark energy, postulating a
string-inspired model of {\it quintessential inflation}, the time
and energy scales involved in the gravitational dynamics may
vastly differ. As we have seen, through the construction of
potentials, the scalar potential driving an inflationary phase at
early epoch and a second weak inflationary episode at late times
could be due to a single exponential term or a sum of exponential
terms, but the slopes of the potential of the leading terms could
be very different. In fact, one of the very interesting features
of an exponential potential is that the cosmological evolution
puts stringent constraints, e.g. during big-bang nucleosynthesis,
on the slope of the potential, but its coefficient may not be
tightly constrained. That is, even if we use the potential
$V(\varphi) = V_0\,e^{-\beta\varphi}$, with $\beta\sim {\cal
O}(1-10)$, for explaining both the early and the late time cosmic
accelerations, the magnitude of the coefficient $V_0$ can be
significantly different, hence indicating completely different
time and energy scales.

\section{When matter fields are present}

The above results were found in the absence of radiation and
matter fields. It is thus natural to ask what happens in a more
realistic situation, at late times, when both radiation and matter
evolve together with the field $\varphi$. In such cases a number
of new and interesting background evolutions might be possible.
Also some of the pathological features, like the appearance of
super-luminal scalar modes, may be absent due to non-trivial
scalar-matter couplings.

\subsection{Minimally coupled scalar field}

We shall first consider the case of a minimal coupling between
$\varphi$ and the matter, so $A(\varphi)=1$. Then, equations
(\ref{GB1})-(\ref{GB3}) and (\ref{cons-eqn-minimal}) can be easily
expressed as a system of three independent equations. After a
simple calculation, we find
\begin{eqnarray}
&& \epsilon=-\frac{1}{2} \left(\ln\Omega_b\right)^\prime
-\frac{3}{2} (1+w_b),\label{eq1-minimal} \\
&&  \mu^\prime=3(1+w_b \Omega_b)+2\epsilon+\frac{\gamma
x^2}{2}-y-(2+\epsilon)\mu,\label{eq2-minimal}
\end{eqnarray}
subject to the constraint $\Omega_\varphi+\Omega_{GB}+\Omega_b=1$.
Note that the background equation of state parameter $w_b$ is a
function of $N$, not a constant. One can write $w_b\Omega_b=w_m
\Omega_m^{(d)}+ w_m^{(s)} \Omega_m^{(s)}+ w_r \Omega_r$, so that
$w_m=0$ and $w_r=1/3$ for pressureless dust and radiation,
respectively, while for a relativistic fluid, e.g. hot and warm
dark matter, we have $w_m^{(s)}>0$. In fact, the first equation
above, (\ref{eq1-minimal}), gives the condition for accelerated
expansion, which requires $\epsilon>-1$. There are two ways to
interpret this equation: either it gives $w_{\rm eff}$ in terms of
$w_b$, assuming that the ratio $\Omega_b^\prime/\Omega_b$ is
known, or it provides for fixed $w_b$ the value of $w_{\rm eff}$
that may be related to the ratio $\Omega_b^\prime/\Omega_b$. Of
course, in the absence of kinetic term for $\varphi$ (so $x=0$),
$f(\varphi)$ is uniquely fixed once $V(\varphi)$ is chosen. Such a
construction is however not very physical.

An interesting question to ask is: What is the advantage of
introducing a non-trivial coupling $f(\varphi)$, if it only
modifies the potential or the variable $y$, without modifying the
functional relation between $\epsilon$ and $\Omega_b$? For the
model under consideration without such a coupling there does not
exist a cosmological solution crossing the phantom divide, $w=-1$.


As a very good late time approximation, suppose that the ratio
$\dot{\varphi}/H$ takes an almost constant value, say $\beta$,
rather than $\dot{\varphi}=0$, along with $H \propto
\,e^{-\,\lambda \phi/2}$, where $\phi\equiv
\sqrt{\gamma}\,(\varphi-\varphi_0)/{\beta}=N$. The background
matter density fraction is then given by
\begin{equation}
\Omega_b = \Omega_0\, \exp[\lambda\phi-3(1+w_b)\phi],
\end{equation}
where $\Omega_0\equiv c_1\, (6w_b + 8-\lambda)/[3(2-\lambda)]$.
For $\lambda<2$, the physical choice is $c_1>0$, so that $\Omega_b
> 0$. The scalar potential and the GB density fraction evolve as
\begin{eqnarray}
V &=&
\frac{3(2-\lambda)+\beta^2(5-\lambda/2)}{2+\lambda}\,e^{-\,\lambda
\phi}- c_1\,e^{-\, 3(1+w_b)\phi}+ c_2\,e^{\,\lambda \phi
/2},\label{pot-with-matter}\\
\Omega_{GB} &\equiv & \mu = \frac{2(\lambda-\beta^2)}{2+\lambda}-
\frac{c_1 (1+w_b)}{2-\lambda}\, e^{\lambda\phi-3(1+w_b)\phi} -
\frac{c_2}{3} \,e^{(2 + \lambda)\phi/2},\label{GB-density-frac}
\end{eqnarray}
where $c_1, c_2$ are integration constants. The second term in
each expression indicates a non-trivial contribution of the
background matter and radiation. The presence of $c_1$ term in
(\ref{pot-with-matter}) implies that the value of the density
fraction $\Omega_b$ must be known rather precisely for an accurate
reconstruction of the potential. Equation (\ref{GB-density-frac})
implies that
\begin{equation}
f(\varphi)\equiv  f_0\,\e^{\lambda \phi} +
f_1\,e^{2\lambda\phi-3(1+w_b)\phi}- f_2\,e^{(2+\lambda/2) \phi},
\end{equation}
where $f_1$ and $f_2$ take, respectively, the signs of $c_1$ and
$c_2$. If $V$ is sufficiently flat near the current value of
$\varphi$, then both the background and Gauss-Bonnet density
fractions, $\Omega_b$ and $\Omega_{GB}$, may take nearly constant
values. At late times, $w_b\simeq w_m \simeq 0$ for the
pressureless dust, while the radiation contribution may be
neglected since $\Omega_r\sim 10^{-4}$.

\subsection{Non-minimally coupled scalar field}

To investigate a possible post-inflation scenario, where the
scalar field may couple non-minimally to a relativistic fluid or
stiff-matter other than to ordinary matter or dust, we consider
the following three different epochs: (1) background domination by
a stiff relativistic fluid, (2) radiation domination and (3) a
relatively long period of dust-like matter domination which
occurred just before the current epoch of dark energy or scalar
field domination. To this end, we define the fractional densities
as follows: $\Omega_s$ for stiff-matter, $\Omega_r$ for radiation
and $\Omega_d$ for dust, where the respective equation of state
parameters are given by $w_s = 1$, $w_r=1/3$ and $w_d = 0$.
Equations (\ref{Q-non-zero-1})-(\ref{Q-non-zero-3}) may be written
as
\bea
\Omega_s^\prime+2\epsilon \Omega_s + 3\Omega_s (1+w_s)&=& - 3
\eta_s Q_s \Omega_s \varphi^\prime, \label{s-continuity} \\
\Omega_d^\prime+2\epsilon \Omega_d +3 \Omega_d &=& - 3 Q_d
\Omega_d
\varphi^\prime.\label{d-continuity} \\
\Omega_r^\prime+2\epsilon \Omega_r +4 \Omega_r &=& 0,
\label{r-continuity}
\eea
In order for current experimental limits on verification of the
equivalence principle to be satisfied, the coupling must be small,
$Q_d^2 \ll1 $, at present, see e.g.~\cite{Damour:2002a}.

Superstring theory in its Hagedorn phase (a hot gas of strings),
and also some brane models, naturally predict a universe filled
with radiation and stiff-matter. To this end, it is not
unreasonable to expect a non-trivial coupling of the field
$\varphi$ with the stiff-matter: highly relativistic fluids may
have strong couplings of the order of unity, $Q_s\sim {\cal
O}(1)$. More specifically, as we do not observe a highly
relativistic stiff-matter at the present time, i.e.
$\Omega_s\approx 0$, the scalar-stiff-matter coupling $Q_s$ in the
order of unity is not ruled out and may have a quantifiable effect
on the evolution of the early universe.

\subsection{The $ \Omega_{GB} =$ const solution}

It is interesting to study the case of a constant $\Omega_{GB}$ as
it allows us to evolve the potential without constraining it
through an ansatz. An exponential potential is generally used in
the literature due to the simplification it affords allowing a
change of variables and hence a system of autonomous equations.

It is worthwhile to investigate whether a physically interesting
result can be found without the need for an ansatz for the
potential. To this end, let us take $\Omega_{GB} \simeq {\rm
const}\equiv \mu_0$. With the ansatz (\ref{specific-coupling}), we
have
\begin{equation}
\frac{{\rm d} x}{{\rm d} N} = -2 x (\epsilon + {x \over
\varphi_0}) \label{uxk-dx}
\end{equation}
\begin{equation}
y = 3 - {\gamma \over 2} x^2 - 3\mu_0 - 3 (\Omega_r + \Omega_s
+\Omega_d), \label{uxk-y}
\end{equation}
where
\begin{equation}
\epsilon = \frac{\gamma x^2 + \mu_0 + 2 x(\eta_s Q_s \Omega_s +
\eta_d Q_d \Omega_d ) + 3 \Omega_r (1 + w_r) + 3 \Omega_s(1 + w_s)
+ 3 \Omega_d (1 + w_d)}{\mu_0 - 2}, \label{uxk-eps}
\end{equation}
in addition to the continuity equations
(\ref{s-continuity})-(\ref{r-continuity}) describing the system.
We also note that, since $\epsilon = H^\prime / H$, $H$ is solved
from equation (\ref{uxk-eps}), allowing us to find both the scalar
field potential $V = y H^2$ and the effective potential
$\Lambda(\varphi)\equiv V(\varphi) + 3 f(\varphi) H^4 (1 +
\epsilon)$ explicitly. We can see that there is a strong
correlation between the length of dust-like matter domination and
the parameter values, $\mu_0$ and $\Omega_{d,\rm ini}$.

\begin{figure}[ht]
\begin{center}
\hskip-0.3cm \epsfig{figure=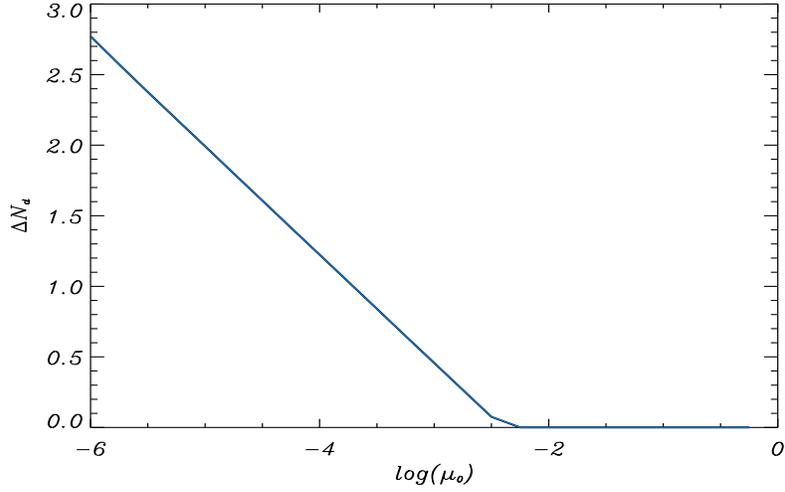,height=2.8in,width=4.5in}
\end{center}
\caption{The period of dust-like matter domination as a function
of $\mu_0$ for $f_{,\varphi}= f_0\,e^{3\varphi}$, $\gamma = 1$,
$Q^2_d = 10^{-5}$ and $Q^2_s = 0.01$. The initial conditions are
$x = 10^{-7}$, $y = 10^{-15}$, $\Omega_d = 0.1$ and $\Omega_r =
0.45$.}
\label{uxkkvl}
\end{figure}

\begin{figure}[ht]
\begin{center}
\hskip-0.3cm \epsfig{figure=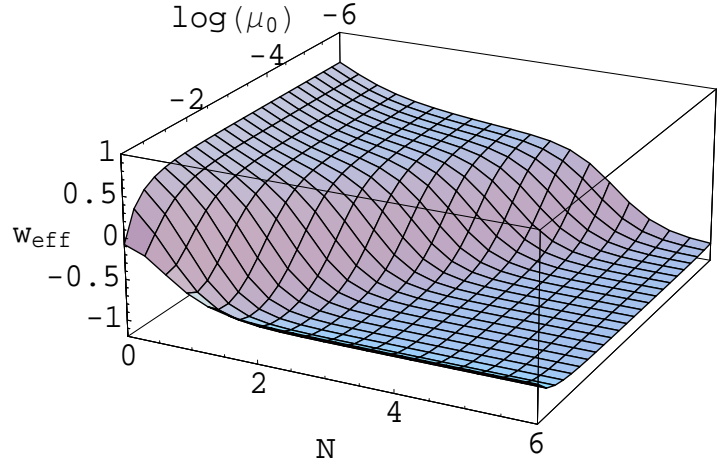,height=2.5in,width=3.8in}
\end{center}
\caption{The behaviour of $w_{\rm eff}$ for varying $\mu_0$ for
$f_{,\varphi}= f_0\,e^{3\varphi}$, $\gamma = 1$, $Q^2_d = 10^{-5}$
and $Q^2_s = 0.01$. The initial conditions are $x = 10^{-7}$, $y =
10^{-15}$, $\Omega_d = 0.1$ and $\Omega_r = 0.45$.} \label{uxkwk}
\end{figure}

The dust domination period is expected to last for about $5-7$
e-folds of expansion. However, in the above case, no parameter
ranges meet this condition while still maintaining a (relatively)
long period of radiation domination. This is problematic as at
least some period of radiation domination appears to be a
requirement for reconciling any cosmological model with
observations. Our results here are therefore presented in the
spirit of a toy model, which allow the evolution of an
unconstrained potential for parameter ranges and give solutions
with the expected qualitative features.

\begin{figure}[ht]
\begin{center}
\hskip-0.2in \epsfig{figure=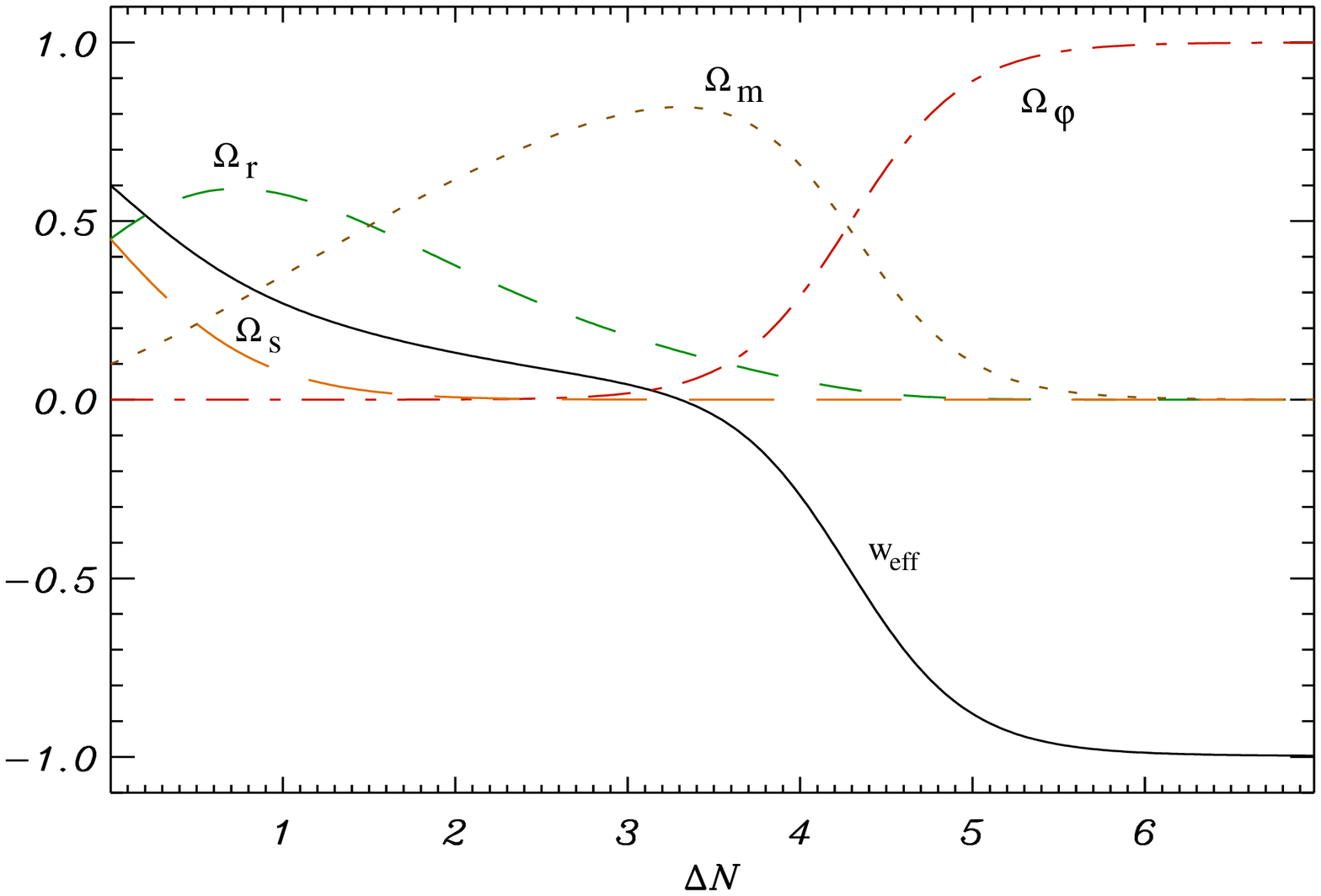,height=2.8in,width=4.5in}
\end{center}
\caption{The evolution of the fractional densities and $w_{\rm
eff}$ (solid line, black) for a fixed $\mu_0 = 10^{-6}$ with
parameters $\gamma=1$, $\varphi_0=2/3$, $Q^2_d = 10^{-5}$ and
$Q^2_s = 0.01$, and initial conditions as $x = 10^{-7}$, $y =
10^{-15}$, $\Omega_d = 0.1$ (dots, brown) and $\Omega_r = 0.45$
(dashes, green). $\Omega_s$ is represented by long dashes (orange)
and $\Omega_\varphi$ by dot-dash (red).} \label{uxktot1}
\end{figure}

\begin{figure}[ht]
\begin{center}
\hskip-0.2cm \epsfig{figure=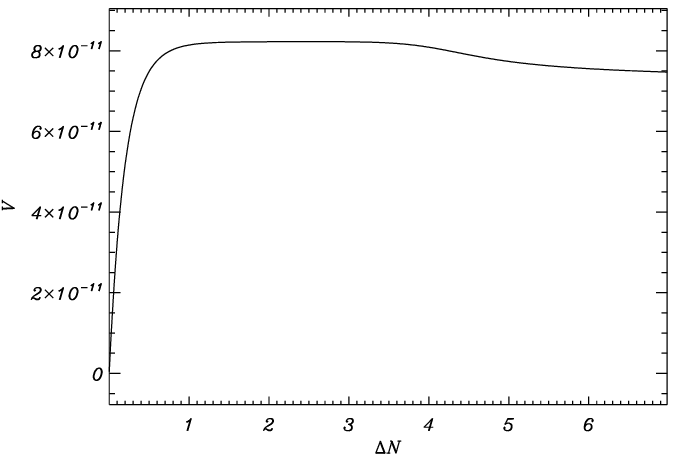,height=2.2in,width=3.0in}
\hskip0.2cm
\epsfig{figure=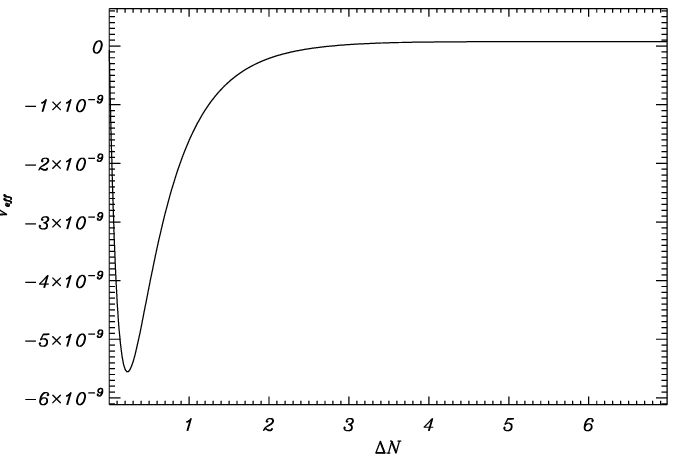,height=2.2in,width=3.0in}
\end{center}
\caption{The left panel shows the evolution of $V(\varphi)$ while
the right panel shows the evolution of $V_{\rm eff} \equiv
V(\varphi) + 3 f(\varphi) H^4 (1 + \epsilon)$ for $\mu_0=10^{-6}$,
$\gamma=1$, $f(\varphi)\propto e^{3\varphi}$, $Q^2_d = 10^{-5}$
and $Q^2_s = 0.01$ with initial conditions $x = 10^{-7}$, $y =
10^{-15}$, $\Omega_d = 0.1$, $\Omega_r = 0.45$, $H = 0.01$ and
$f(\varphi)\Z{\rm ini} = 10^{-10}$.} \label{uxkv2}
\end{figure}

Figure~\ref{uxkkvl} shows the period of dust domination observed
for various values of $\mu_0$. We see that, in the parameter range
$ \mu_0 \lesssim 10^{-3}$, there exist solutions supporting both
radiation and stiff-matter domination prior to matter domination.
While for larger values of $\mu_0$ the Gauss-Bonnet contribution
suppresses the background fields and we may observe a quick
transition to scalar-field domination, missing either the
radiation dominated era or the matter dominated era, or both. For
smaller values of $\mu_0$, no solutions exist for the parameters
used above due to constraints in the field equations. The
behaviour of $w_{\rm eff}$ for varying $\mu_0$ and the general
evolution of the cosmology are shown in figure~\ref{uxkwk} and
figure~\ref{uxktot1}, respectively. In the cases where the second
derivative of $w_{\rm eff}$ is monotonic there is no radiation
domination ($w_{\rm eff}\sim 1/3$) or dust domination ($w_{\rm
eff}\sim 0$). For lower values of $\mu_0$, the second derivative
of $w_{\rm eff}$ obviously changes sign and hence implies a period
of background matter or radiation domination. We show the
behaviour of the scalar potential and the corresponding effective
potential in figure~\ref{uxkv2}.

\subsection{Simplest exponential potentials}

We now wish to consider the evolution of the full system while
putting minimal restrictions on the evolution of the cosmological
constituents. To do this we employ simple single exponential terms
for both the field potential and the scalar-GB coupling:
\begin{equation}\label{f-v-ansatz}
f_{, \varphi} (\varphi) = f_0\, e^{\alpha \varphi} \hspace{0.6cm}
{\rm and} \hspace{0.6cm} V(\varphi) = V_0\, e^{-\beta \varphi}.
\label{ansatz-f,v}
\end{equation}
The commonly invoked exponential potential ansatz has some
physical motivation in supergravity and superstring theories as it
could arise due to some nonperturbative effects, such as gaugino
condensation and instantons. One can see that these choices are
almost undoubtedly too naive to allow all the expected physical
features of our universe from inflation to the present day. This
is because generally the slopes of the potential considered in
post-inflation scenarios are too steep to allow the required
number of e-folds of inflation in the early universe. As a
post-inflation approximation, however, these may hold some
validity, as one can replicate many observable physical features
from nucleosynthesis to the present epoch while allowing
non-trivial scalar-matter couplings. In follow-up work we will
discuss the possibility of a two-scalar fields model where the
potential related to one scalar field meets the requirements of
inflation while the other scalar drives the late time cosmology.

The ans\"atze (\ref{f-v-ansatz}) allow us to write an autonomous
system
\bea
\quad {{\rm d} x \over {\rm d} N} &=& - {1 \over 2 \gamma} \big[ 2
\gamma x \epsilon + 6 \gamma x - 2 \beta y + 6 u (1 + \epsilon) -
6 \eta_s
Q_s \Omega_s  \nonumber \\
& & \quad + \eta_d Q_d \{ \gamma x^2 + 2 y
+ 6( u x + \Omega_r + \Omega_s -1 ) \} \big] \\
{{\rm d} y \over {\rm d} N} &=& - y ( \beta x + 2 \epsilon ) \\
{{\rm d} u \over {\rm d} N} &=& u ( \alpha x + 2 \epsilon )
\eea
where
\bea
\hspace{0.5cm} \epsilon &=&  \frac{1}{6(2 \gamma u x - 2 \gamma -
3 u^2)} \bigg[ 18 \gamma \left( 1 + w_d - w_d
\Omega_r - w_d \Omega_s + w_r \Omega_r + w_s \Omega_s \right) \nonumber \\
& & + 6 \gamma u x - 6 \gamma y + 3 \gamma^2 x^2 + 18 u^2 - 3
\gamma^2 w_d x^2 - 6 \beta u y - 6 \gamma \alpha u x^2 - 18 \gamma
w_d u x - 6 \gamma w_d y \nonumber \\ & & - \eta_d Q_d ( 2
\gamma^2 x^3 + 18 u - 18 u^2 x - 6 u y - 18 u \Omega_r - 12 \gamma
x + 4 \gamma x y + 12 \gamma x \Omega_r + 9 \gamma u x^2 )
\nonumber \\ & & - \left( \eta_d Q_d - \eta_s Q_s \right) \left(
12 \gamma x - 18 u \Omega_s \right) \bigg],
\eea
which along with the continuity equations
(\ref{s-continuity})-(\ref{r-continuity}) and the Friedmann
constraint equation, $\Omega_d + \Omega_\varphi + \Omega_{GB} +
\Omega_r + \Omega_s=1$, allow us to proceed with numerical
computation.

\begin{figure}[ht]
\begin{center}
\hskip-0.5cm
\epsfig{figure=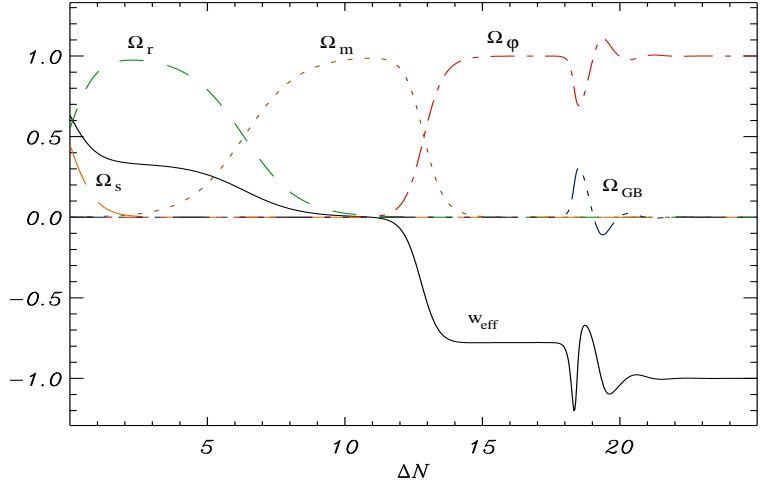,height=2.8in,width=4.5in}
\end{center}
\caption{Evolution of the fractional densities and $w_{\rm eff}$
(solid line, black) where $\beta = \sqrt{2/3}$, $\gamma=1$,
$\alpha=12$, $Q_d^2 = {10}^{-5}$ and $Q_s^2 = 0.01$ with initial
conditions $x = \sqrt{6} \times 10^{-4}$, $y = 5 \times 10^{-20}$,
$u = 0.08$, $\Omega_r = 0.549$ (dashes, green) and $\Omega_s =
0.45$ (long dashes, orange). $\Omega_d$ is represented by dots
(brown), $\Omega_\varphi$ by dot-dash (red) and $\Omega_{GB}$ by
dot-dot-dot-dash (blue).} \label{full-totsq23}
\end{figure}

\begin{figure}[ht]
\begin{center}
\hskip-0.5cm
\epsfig{figure=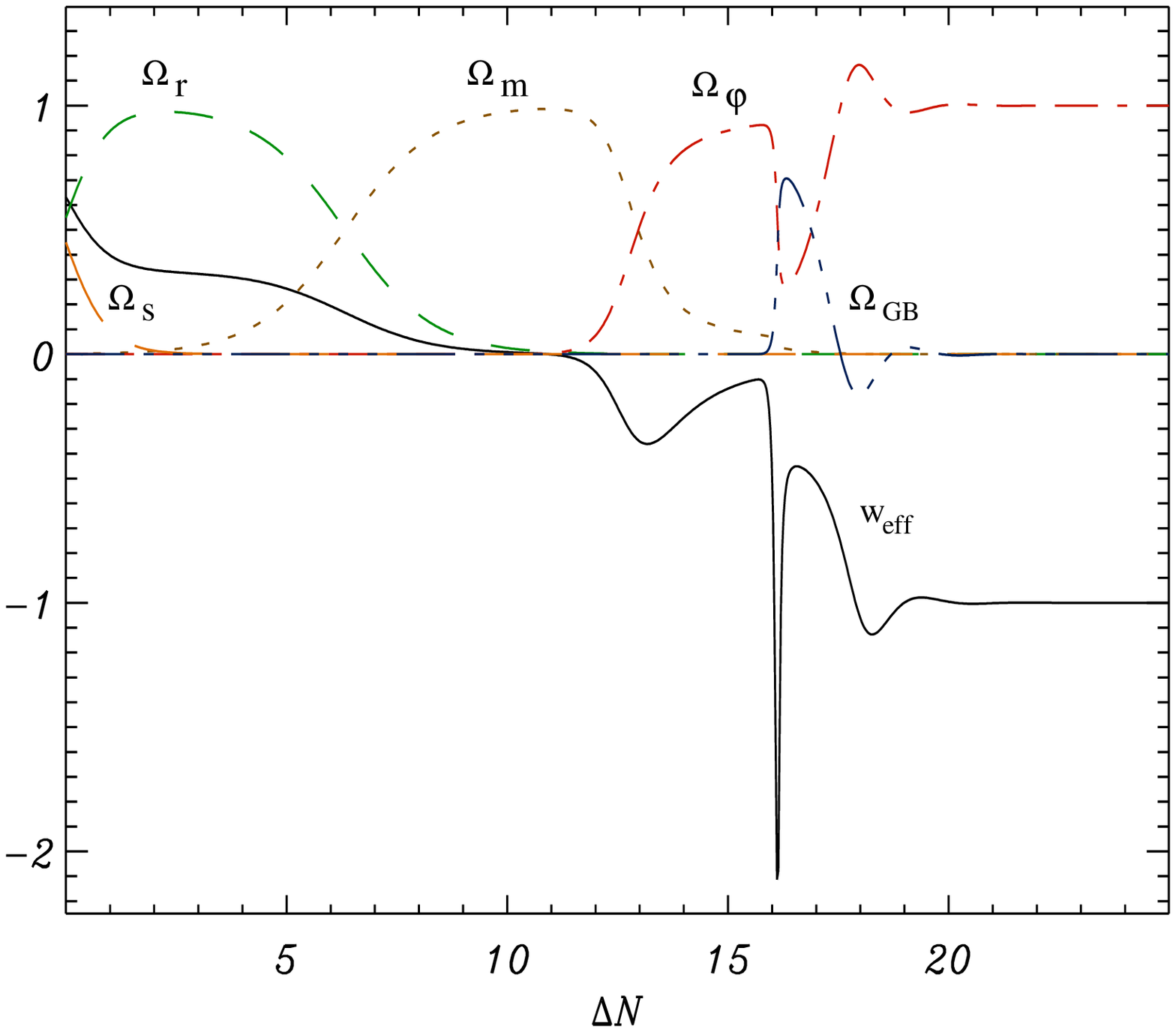,height=2.8in,width=4.5in}
\end{center}
\caption{Evolution of the fractional densities and $w_{\rm eff}$
(solid line, black) where $\beta = \sqrt{3}$, $\gamma=1$,
$\alpha=12$, $Q_d^2 = {10}^{-5}$ and $Q_s^2 = 0.01$ with initial
conditions $x = \sqrt{6} \times 10^{-4}$, $y = 5 \times 10^{-20}$,
$u = 0.08$, $\Omega_r = 0.549$ (dashes, green) and $\Omega_s =
0.45$ (long dashes, orange). $\Omega_d$ is represented by dots
(brown) and $\Omega_\varphi$ by dot-dash (red), $\Omega_{GB}$ by
dot-dot-dot-dash (blue).} \label{full-totsq3}
\end{figure}

We look particularly at two values for $\beta$ here, $\beta =
\sqrt{2/3}$ and $\beta = \sqrt{3}$, while keeping the other
parameters constant to limit the vast parameter space. These
values may be motivated by various schemes of string or M theory
compactifications (see, e.g.~\cite{Neupane:2005nb}). We will
discuss the effects of varying these other parameters and some of
the physically relevant results. The evolution of the various
constituents is shown in
figures~\ref{full-totsq23}-\ref{full-totsq3}.

The $\beta = \sqrt{2/3}$ case appears to show a smoother evolution
and has a short period during which there is acceleration and an
appreciable amount of matter in the universe. In the table below
we look at some of the features of the accelerating period for the
solution given in figure~\ref{full-totsq23}. At the onset of
acceleration, $w_{\rm eff} = -1/3$, we have $\Omega_d = 0.650$.
Within the best-fit concordance cosmology the present
observational data seem to require $w_{\rm
eff}<-0.74$~\cite{Wang04py}. However, in the above case, this
requires less matter ($\Omega_d\lesssim 0.12$) than in
$\Lambda${CDM} model.

\vspace{0.3cm}
\begin{center}
\begin{tabular}{|c|c|c|c|}
  \hline
  Condition imposed at present
  & $\Delta N_{\rm accelerating}$ & Implied $w_{\rm eff}$ & Implied $\Omega_d$ \\
  \hline
  $\Omega_d = 0.27$ & $0.582$ & $-\, 0.648$ & N/A\\
  $w_{\rm eff}= -\, 0.74$& $0.955$ & N/A & $0.123$\\
  $w_{\rm eff}=-\, 0.9$& $5.48$ & N/A & $3.02 \times 10^{-6}$\\
  $w_{\rm eff}=-\, 1.0$& $5.55$ & N/A & $2.46 \times 10^{-6}$\\
  $\Delta N = 0.69$ & N/A & $-\, 0.685$ & $0.215$\\
  $\Delta N = 0.91$ & N/A & $-\, 0.734$ & $0.134$\\
  \hline
\end{tabular}
\end{center}

\vspace{0.3cm}
The recent type Ia supernovae observations~\cite{Riess:2006}
appear to indicate that the universe may be accelerating out to a
redshift of $z \sim 0.4 - 1$~\cite{Alam:2006}. In terms of the
number of e-folds of expansion, this corresponds to $\Delta N=\ln
(1+z) \sim 0.34-0.69$, if one assumes a rescaling of $N
\rightarrow 0$ at the present epoch using the freedom in choosing
the initial value of the scale factor, $a_0$. We have chosen the
initial value of $y$, or the ratio $V(\varphi)/H^2$, such that the
period of dust-like matter domination is $\Delta N_{\Omega_d} \sim
6.5$ as this corresponds to a total redshift of $z \sim 1100$ to
the epoch of
matter-radiation equality. 

\begin{figure}[ht]
\begin{center}
\hskip-0.5cm
\epsfig{figure=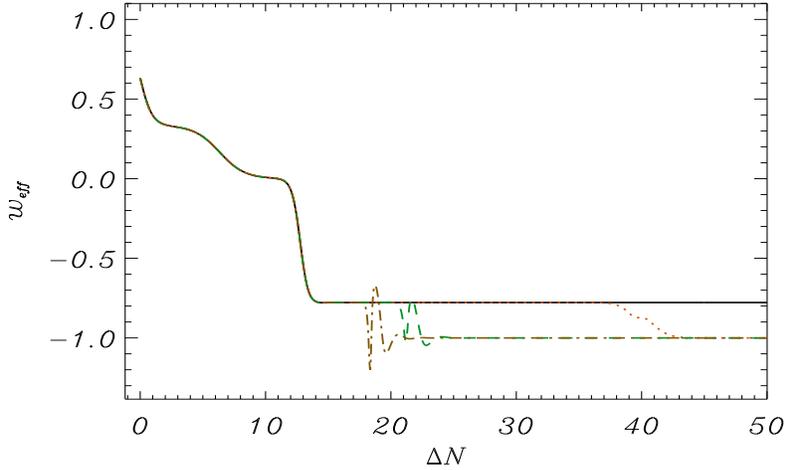,height=2.8in,width=4.5in}
\end{center}
\caption{Evolution of $w_{\rm eff}$ with $\beta=\sqrt{2/3}$,
$\gamma=1$, $Q_d^2 = 10^{-5}$ and $Q_s^2 = 0.01$; $\alpha = 12$
(dash-dot, brown), $8$ (dashes, green), $3$ (dots, red),
$\sqrt{2/3}$ (solid, black), with initial values $x = \sqrt{6}
\times 10^{-4}$, $y = 5 \times 10^{-20}$, $u = 0.08$, $\Omega_r =
0.549$ and $\Omega_s = 0.45$.} \label{alwsq23}
\end{figure}

\begin{figure}[ht]
\begin{center}
\hskip-0.5cm \epsfig{figure=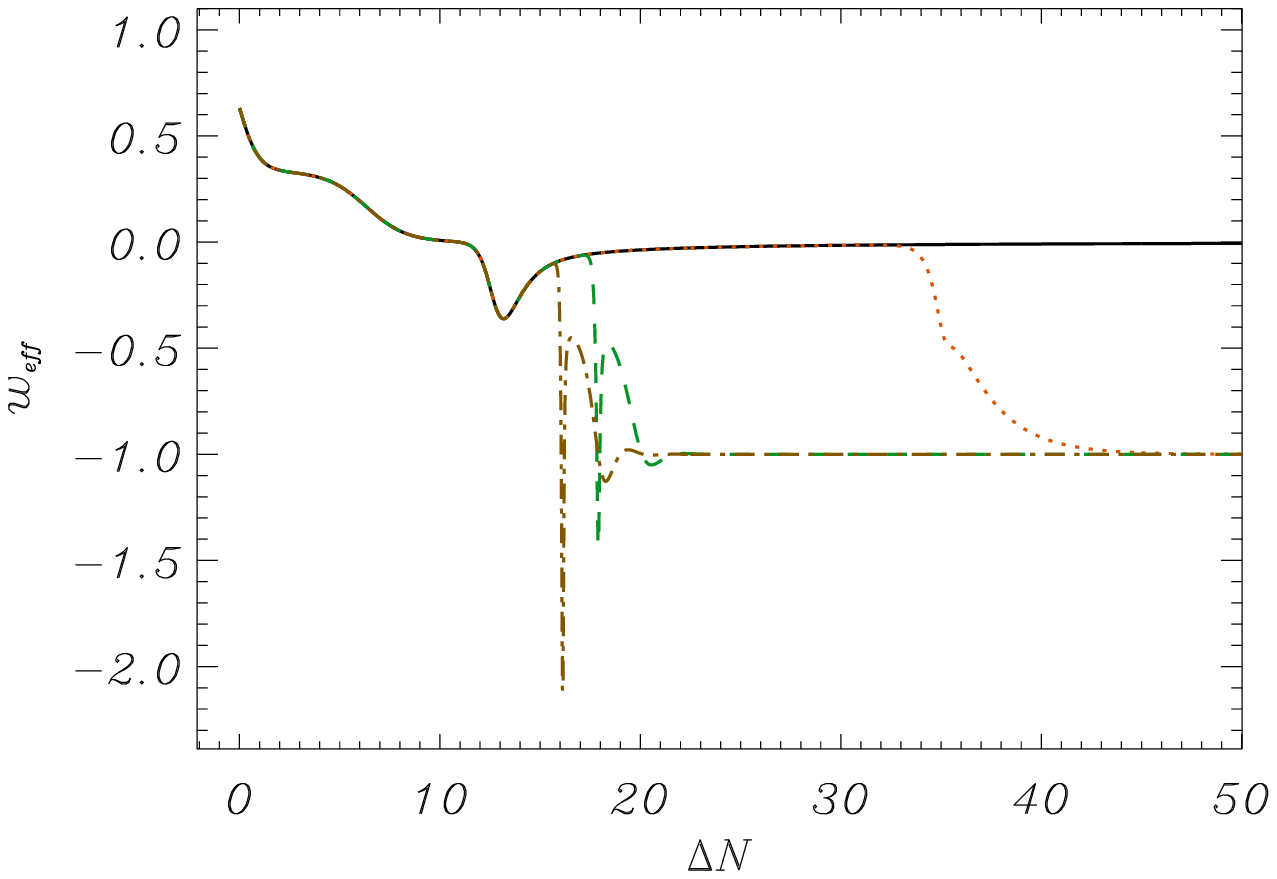,height=3in,width=4.5in}
\end{center}
\caption{Evolution of ${\rm w}_{\rm eff}$ with $\beta = \sqrt{3}$,
$\gamma=1$, $Q_d^2 = 10^{-5}$ and $Q_s^2 = 0.01$; $\alpha = 12$
(dash-dot, brown), $8$ (dashes, green), $3$ (dots, red),
$\sqrt{2/3}$ (solid, black), with initial values $x = \sqrt{6}
\times 10^{-4}$, $y = 5 \times 10^{-20}$, $u = 0.08$, $\Omega_r =
0.549$ and $\Omega_s = 0.45$.} \label{alwsq3}
\end{figure}

\begin{figure}[ht]
\begin{center}
\epsfig{figure=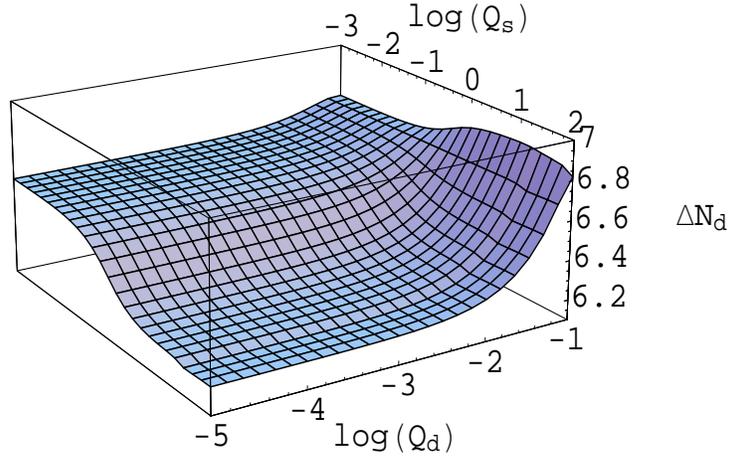,height=2.5in,width=3.8in}
\end{center}
\caption{The period of dust-like matter domination for varying
$Q_s$ and $Q_d$ for $\gamma=1$, $\alpha=12$ and $\beta=\sqrt{2/3}$
with initial conditions $x = \sqrt{6} \times 10^{-5}$, $y = 5
\times 10^{-20}$, $u = 0.08$, $\Omega_r = 0.549$ and $\Omega_s =
0.45$.} \label{qv}
\end{figure}

\begin{figure}[ht]
\begin{center}
\hskip-0.3cm \epsfig{figure=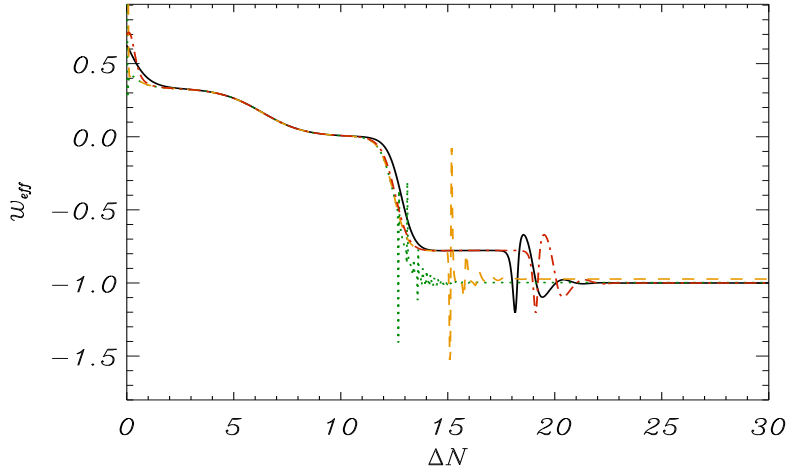,height=2.8in,width=4.5in}
\end{center}
\caption{Evolution of $w_{\rm eff}$ for different values of $Q_s$:
$Q_s^2 = 10^4$ (dots, green) $100$ (dashes, orange), $0$ (solid,
black) and $1$ (dash-dot, red) with fixed $Q_d=\sqrt{10^{-5}}$ and
$\gamma=1$, $\alpha=12$, $\beta=\sqrt{2/3}$. The initial
conditions are $x = \sqrt{6} \times 10^{-4}$, $y = 5 \times
10^{-20}$, $u = 0.1$, $\Omega_r = 0.549$ and $\Omega_s = 0.45$.
The larger, more violent oscillations seen for $Q_s > 1$ are due
to significant scalar-stiff-matter couplings, while the smooth
oscillations for smaller couplings at some later stage are due to
an appreciable contribution of the coupled Gauss-Bonnet term.}
\label{qw1}
\end{figure}
\begin{figure}[ht]
\begin{center}
\hskip-0.2cm \epsfig{figure=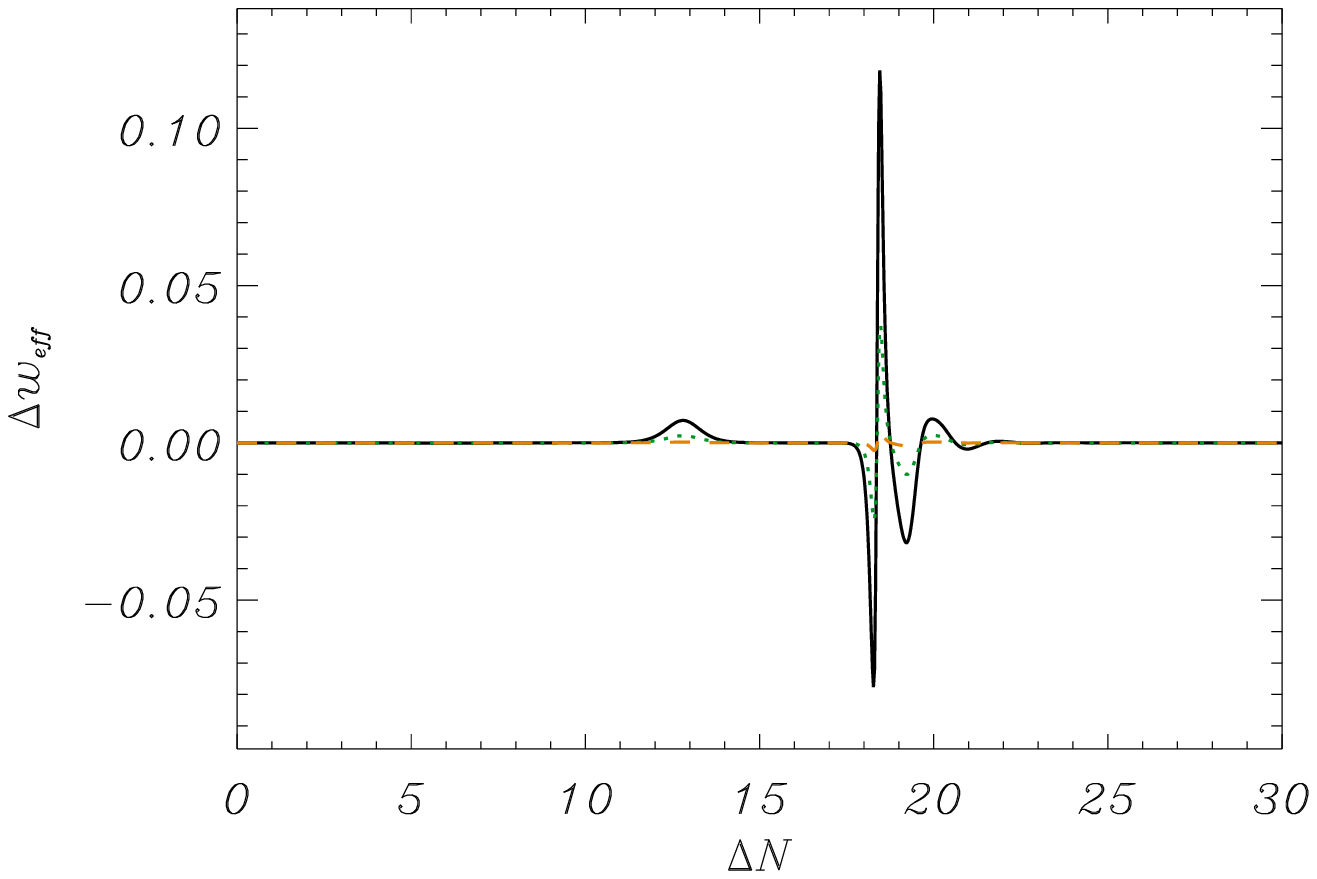,height=2.8in,width=4.5in}
\end{center}
\caption{Variation of $w_{\rm eff}$ for non-zero $Q_d$ as compared
to the $Q_d=0$ case, $\Delta w_{\rm eff}\equiv w_{\rm eff}(Q_d\neq
0)-w_{\rm eff}(Q_d=0)$; $Q_d^2 = 10^{-5}$ (solid, black),
$10^{-6}$ (dots, green) and $10^{-8}$ (dashes, orange) and
$Q_s=0.1$. Clearly, for higher values of the scalar dust-matter
couplings, $Q_d^2> {10}^{-5}$, we find slightly larger variations
in $w_{\rm eff}$.} \label{qw2}
\end{figure}

The evolution with $\beta = \sqrt{3}$ does not seem to be in
favourable agreement with current cosmological observations. The
solution may undergo a period of sudden change in $w_{\rm eff}$
when the Gauss-Bonnet contribution becomes appreciable (or
significant); this would have to occur around at the present epoch
as to retrieve the current value of $w_{eff}\sim -1$. This result
does not reconcile with either the supernovae data, which seem to
indicate a longer period of acceleration, or with a constraint for
the present value of Gauss-Bonnet density~\cite{Amendola:2005}
which may require that $\Omega_{GB}\lesssim 0.2$, or with the
presently anticipated value of $\Omega_d$ ($\sim 0.27$).

We can observe an oscillatory crossing of $w_{\rm eff}= -1$ limit
for all cases in which the Gauss-Bonnet contribution becomes
appreciable, even momentarily. Such a behaviour may be seen in
variants of scalar-tensor models~\cite{phan-cross-obs}. In our
case, the amplitude of these oscillations corresponds to the
amplitude of the oscillations seen in the Gauss-Bonnet
contribution and hence is heavily dependent on the slope of the
scalar-GB coupling; for large $\alpha$ we observe much larger
oscillations. These oscillations damp quickly as the Gauss-Bonnet
contribution becomes negligibly small, and settle to a late time
evolution for which $w_{\rm eff} \approx -1$. As this limit is
approached from above, none of the issues inherent with
super-inflation or a violation of unitarity will be applicable to
the late time cosmology.

As the parameter space for the initial conditions is very large we
have presented solutions with initial conditions and parameters
selected to give reasonable periods of radiation and dust-like
matter domination as well as other physically favourable features.
Although the quantitative behaviour is observed to change smoothly
with changes in initial conditions and parameters, such changes do
have some qualitative effects as limits of certain behaviour are
encountered. A quantitative variation in the period of dust-like
matter domination can be attributed to altering the initial value
of $y$ or $\Omega_{d,\rm ini}$; lower values of $y_{\rm ini}$
extend the period before scalar field domination begins. This is
effectively a change in the initial potential and has a monotonic
effect on the epoch at which scalar field domination begins. In a
qualitative sense, it has no effect on the period of radiation
domination until a $y_{\rm ini}$ is selected which is large enough
that the scalar field contribution completely suppresses the
dust-like matter domination period. The value of $\Omega_{d,\rm
ini}$ is relevant to the epoch of matter-radiation equality,
larger values result in an earlier epoch. For small values, no
dust-like matter domination occurs and the solution is entirely
dominated by the other four constituents considered. The limits at
which all these effects occur are dependent on other parameters
and hence discussion of actual values instead of the general
behaviour does not add further insight.

We note that the exponential terms for both the potential and the
coupling have some noticeable effects on the evolution of the
system. This may be seen to an extent in terms of $\beta$ in
figures~\ref{full-totsq23}-\ref{alwsq3}. It does, however, appear
that the ratio of these parameters also influences the expansion
during dust-like matter domination remaining, relatively unchanged
when this ratio is constant within realistic $\beta$ and $\alpha$
parameter ranges. Phenomenological bounds on these values have
been studied in \cite{Koivisto:2006}.

For $\alpha \sim \beta$ we do not observe a significant period of
Gauss-Bonnet contribution and hence no crossing of the $w_{\rm
eff}= -1$ limit. The closer the value of $\alpha$ is to this limit
the later this period of significant Gauss-Bonnet density fraction
occurs. As $\alpha$ increases there is a minimum epoch at which
the Gauss-Bonnet contribution becomes significant, this epoch
occurs after the scalar field becomes the dominant component in
the energy budget of our universe. Results for various values of
$\alpha$ are shown in figures \ref{alwsq23} and \ref{alwsq3}.

The consequences of matter coupling to the scalar field on the
cosmology are of particular interest. The effect in the period of
dust-like matter-domination is minimal in relation to both $Q_d$
and $Q_s$ and can be seen in figure~\ref{qv}, though we allowed
the value $Q_s\gg Q_d$. As there is a lack of observational or
theoretical motivation refuting the possibility of a high
relativistic-matter-scalar coupling we consider a range that
extends beyond $Q_s = 1$ whereas the dust-like-matter-scalar
coupling must take values $Q_d^2 < 10^{-5}$ due to the current
level of experimental verification in solar system tests of GR.

The solution undergoes a transition in qualitative behaviour when
we extend through the limit $Q_{s, \rm lim} \sim 1$, where the
exact onset and amplitude of its effects are dependent on the
other parameter values taken. Hence our remarks again apply to the
general behaviour observed rather than specific cases. Couplings
greater than $Q_{s, \rm lim}$ cause a non-negligible reemergence
of the stiff-matter contribution at the end of dust-like matter
domination. The density fraction of the relativistic matter
undergoes a damped oscillations, generally with a much shorter
period of oscillation than the oscillations observed in the
non-negligible Gauss-Bonnet contribution. This causes
corresponding oscillations in the effective EoS, while still
generally showing the same overall trend as for lower $Q_s$ values
of the otherwise same solution, with the density fraction of the
relativistic matter stabilising to a non-zero late time value.
This cosmological behaviour does not appear to be physically valid
as no mechanism to generate this relativistic matter seems
plausible and would hence lead us to suggest that the value of
$Q_s$ would have an upper-bound such that $Q_{s, \rm max} \sim
Q_{s, \rm lim}$. The effects of $Q_s$ and $Q_d$ on the effective
equation of state parameter, $w_{\rm eff}$, may be seen in
figures~\ref{qw1} and \ref{qw2}. Note that in figure~\ref{qw2} we
have considered $\Delta w_{eff}$ rather than $w_{eff}$ as the
effects of varying $Q_d$ are so minimal that no discernable
variation can be seen otherwise. In these figures we have
considered the same solutions as in figure~\ref{full-totsq23}. The
magnitude of the effects are, however, the same throughout the
parameter space.

\begin{figure}[ht]
\begin{center}
\epsfig{figure=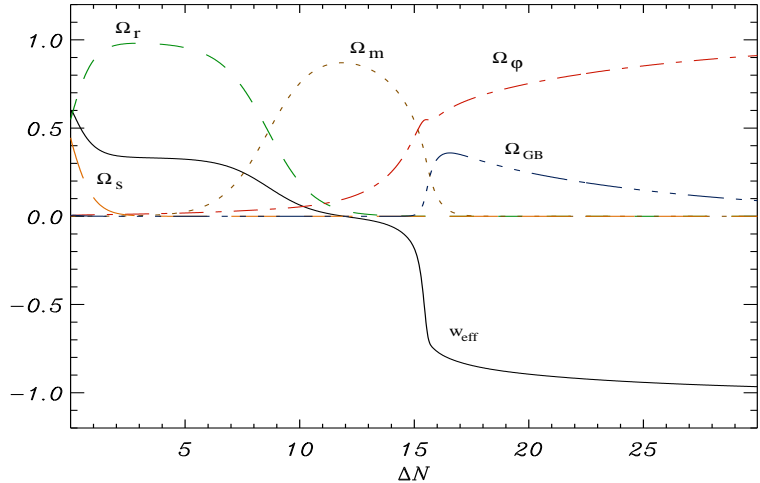,height=2.8in,width=4.5in}
\end{center}
\caption{Evolution of the fractional densities and $w_{\rm eff}$
(solid line, black) for the potential ansatz $V(\varphi) =
H^2(\Lambda_0 + \Lambda_1 e^{-\beta \varphi})$ where $\gamma=1$,
$\alpha=9$, $\beta = \sqrt{2/3}$, $\Lambda_0 = 10^{-8}$, $Q_d^2 =
{10}^{-5}$ and $Q_s^2 = 0.01$ with initial conditions $x =
\sqrt{6} \times 10^{-2}$, $y = 9 \times 10^{-3}$, $u = 0.1$,
$\Omega_r = 0.549$ and $\Omega_d = 10^{-4}$. $\Omega_{\rm r}$ is
represented by dashes (green), $\Omega_{\rm d}$ by dots (brown)
$\Omega_s$ by long dashes (orange), $\Omega_\varphi$ by dot-dash
(red) and $\Omega_{GB}$ by dot-dot-dot-dash (blue).}
\label{altpot}
\end{figure}

\subsection{A canonical potential}

Motivated by our results in section 3, we consider the potential
\begin{equation}
V(\varphi) = H^2(\Lambda_0 + \Lambda_1 e^{\beta \varphi}).
\end{equation}
Imposing this ansatz along with the scalar-GB coupling given in
equation (\ref{ansatz-f,v}), $f(\varphi)\propto
e^{\alpha\varphi}$, allows us to find solutions that have
reasonable agreement with concurrent observations while not
crossing the $w_{\rm eff}= -1$ limit at any stage of the
evolution. From figure~(\ref{altpot}) we can see that the cosmic
evolution shows a smooth progression to $w_{\rm eff}= -1$, which
may be physically more sensible. The amplitude of the Gauss-Bonnet
density fraction at maximum is dependent on the values of slope
$\alpha$, for smaller $\alpha$ it never becomes relevant. The
period of dust-like matter domination again shows a heavy
dependence on the initial value of $y$ or the potential
$V(\varphi)$.

Here we also make an essential remark related to all our numerical
plots in sections $5$ and $6$. Taking into account the relation
$\Delta N\equiv \ln (1+z)$, one may (incorrectly) find that the
present epoch corresponds to $\Delta N\sim 15$. However, in our
numerical analysis, we are interested in the general evolution of
the universe, in terms of the e-folding time $N\equiv \ln[a(t)]$,
from the (early) epoch of radiation-dominated universe to the
present epoch. Indeed, our solutions are completely invariant
under a constant shift in $\Delta N$, so the choice of $N\Z0$ in
$\Delta N= N+ N\Z0$ is arbitrary. One may choose $N\Z0$ such that
$\Omega\Z{m}\simeq 0.27$ at present, or alternatively $N\Z0 \sim
15$, so that $a\Z0=1$ (the present value of the scale factor)
corresponds to $N\simeq 0$.

\section{Remarks on ghost conditions}

\begin{figure}[ht]
\begin{center}
\epsfig{figure=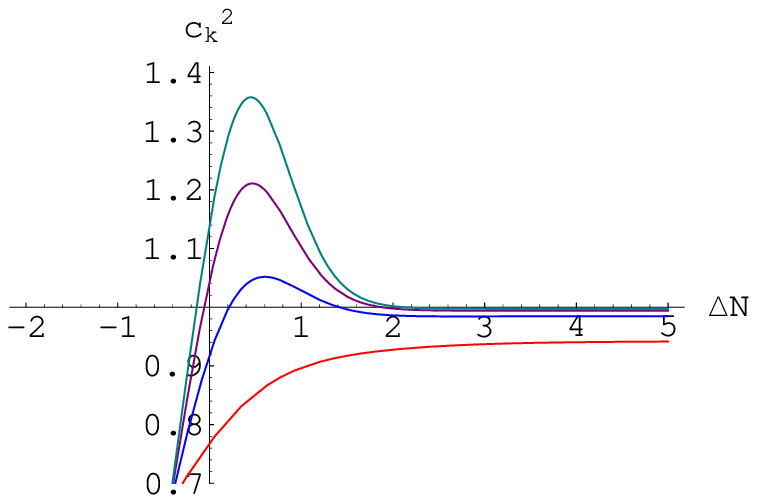,height=2.0in,width=2.8in}
\hskip0.2in
\epsfig{figure=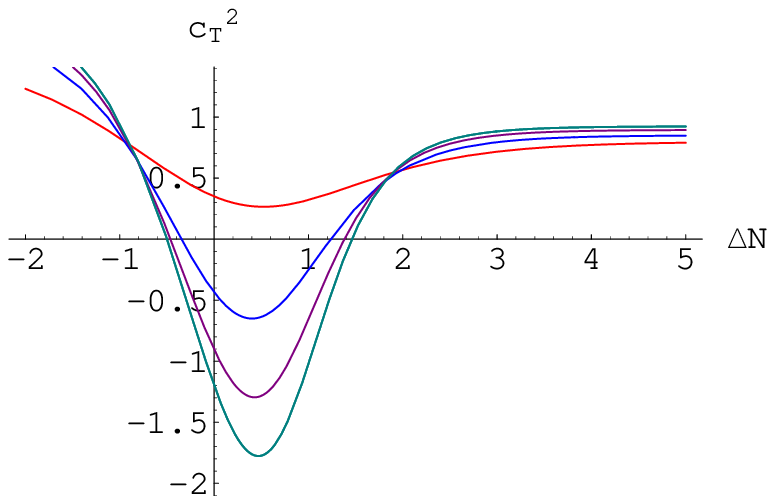,height=2.0in,width=2.8in}
\end{center}
\caption{The speeds of propagation for scalar and tensor modes,
corresponding to the solution (4.21) with $\sqrt{\gamma}
\varphi_0= 3, 4, 5$ and $6$ from top to bottom (bottom to top) for
the left (right) panel.} \label{tensor-speed1}
\end{figure}

\begin{figure}[ht]
\begin{center}
\hskip-0.3cm
\epsfig{figure=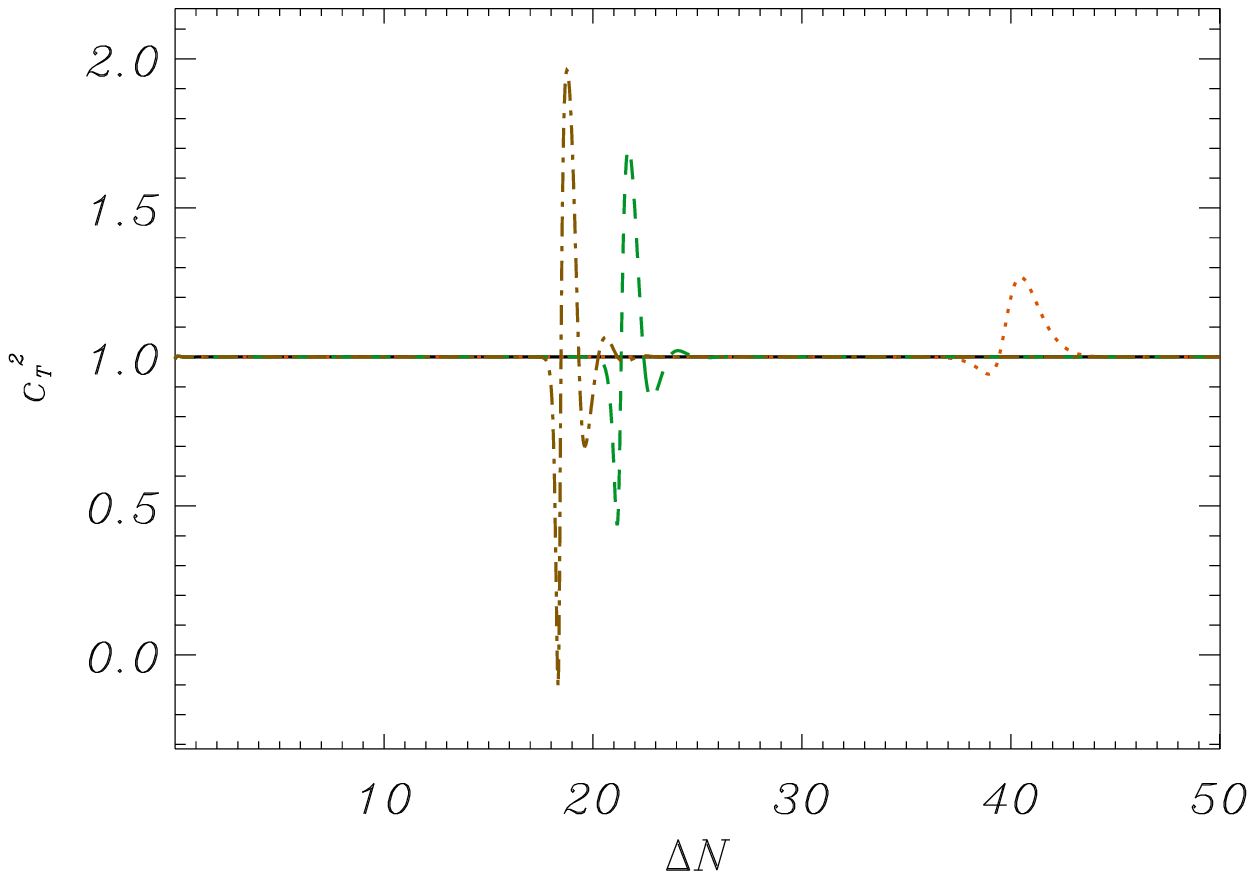,height=2.8in,width=4.5in} \vskip0cm
\epsfig{figure=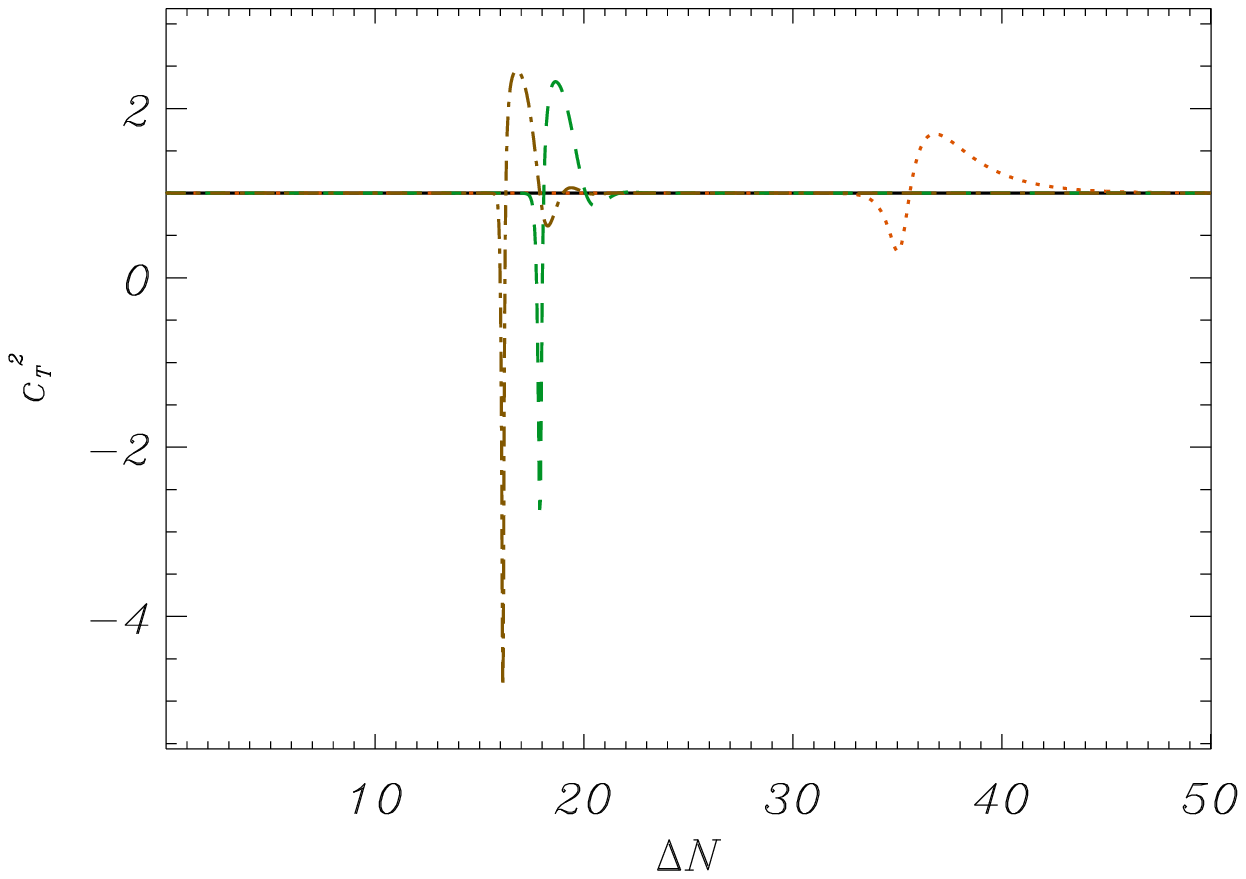,height=2.8in,width=4.5in}
\end{center}
\caption{The variation of $c_T^2$ with $\Delta N$; (top panel)
$\beta = \sqrt{2/3}$ and $\alpha = 12 \,{\rm (dash-dot, brown)},\,
8 \, {\rm (dashes, green)},\, 3 \, {\rm (dots, orange)}, \,
\sqrt{2/3} \, {\rm (solid, black)}$; (bottom panel) $\beta =
\sqrt{3}$ and $\alpha = 12 \,{\rm (dash-dot, brown)},\, 8 \, {\rm
(dashes, green)},\, 3 \, {\rm (dots, orange)}, \, \sqrt{2/3} \,
{\rm (solid, black)}$. Other parameters are chosen as $\gamma=1$,
$Q_d^2 = 10^{-5}$ and $Q_s^2 = 0.01$, with initial conditions $x =
\sqrt{6} \times 10^{-4}$, $y = 5 \times 10^{-20}$, $u = 0.08$,
$\Omega_r = 0.549$ and $\Omega_s = 0.45$.} \label{ct}
\end{figure}

\begin{figure}[ht]
\begin{center}
\hskip-0.3cm
\epsfig{figure=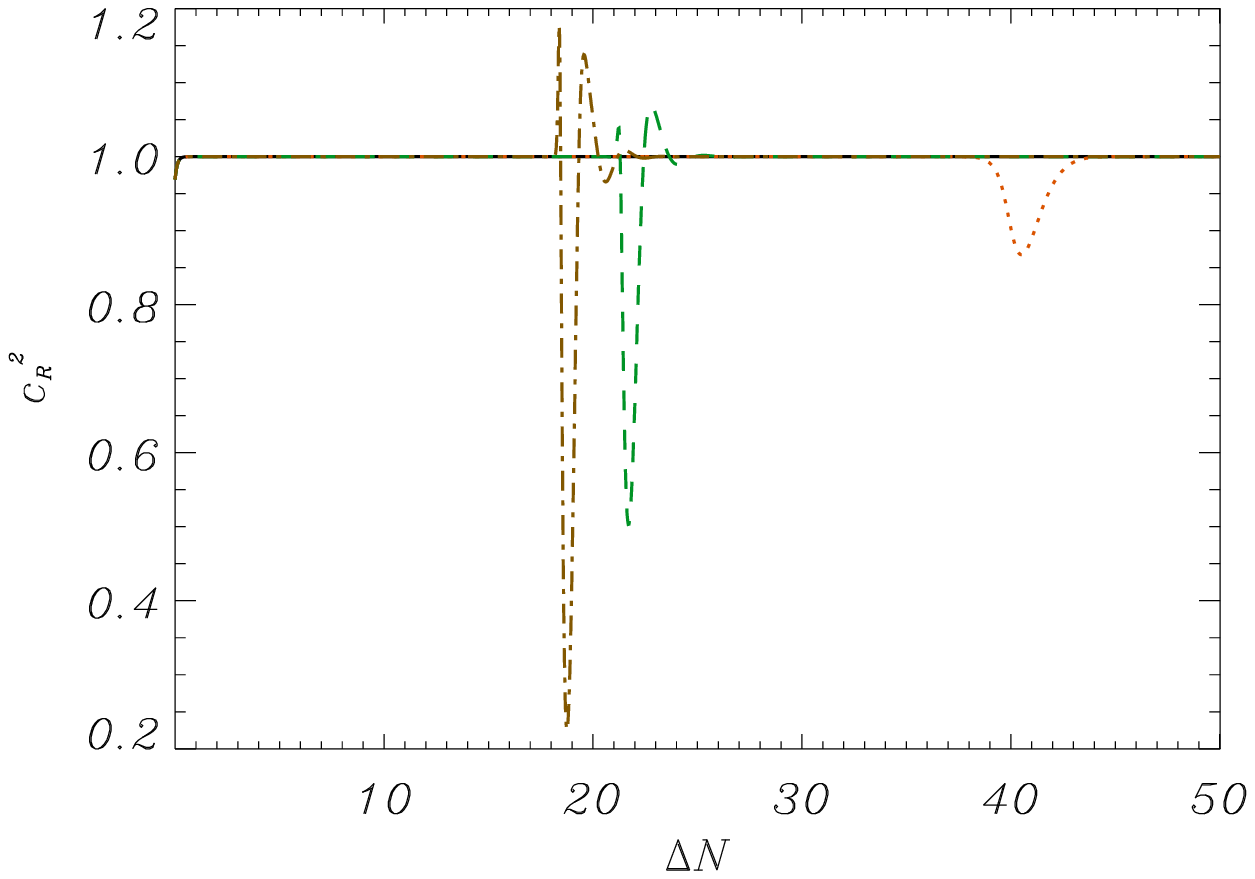,height=2.8in,width=4.5in}
\vskip0.1cm
\epsfig{figure=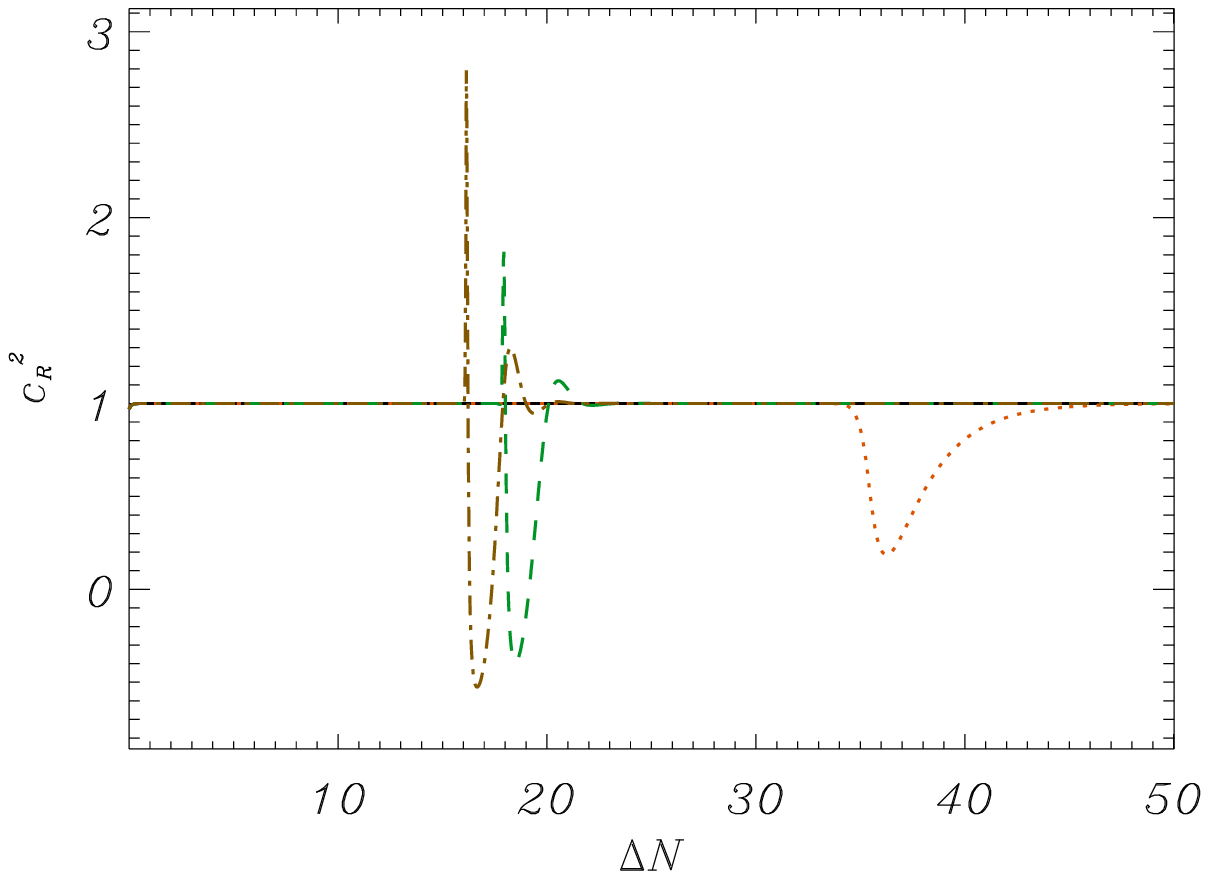,height=2.8in,width=4.5in}
\end{center}
\caption{The variation of $c_{{\cal R}}^2$ with $\Delta N$; (top
panel) $\beta = \sqrt{2/3}$ and $\alpha = 12 \,{\rm (dash-dot,
brown)},\, 8 \, {\rm (dashes, green)},\, 3 \, {\rm (dots,
orange)}, \, \sqrt{2/3} \, {\rm (solid, black)}$; (bottom panel)
$\beta = \sqrt{3}$ and $\alpha = 12 \,{\rm (dash-dot, brown)},\, 8
\, {\rm (dashes, green)},\, 3 \, {\rm (dots, orange)}, \,
\sqrt{2/3} \, {\rm (solid, black)}$, for $\gamma=1$, $Q_d^2 =
10^{-5}$ and $Q_s^2 = 0.01$ with initial conditions $x = \sqrt{6}
\times 10^{-4}$, $y = 5 \times 10^{-20}$, $u = 0.08$, $\Omega_r =
0.549$ and $\Omega_s = 0.45$.} \label{cs}
\end{figure}

Recently, a number of
authors~\cite{Calcagni:06ye,DeFelice:2006pg}~\footnote{The model
studied in~\cite{DeFelice:2006pg} may find only limited
applications within our model, as the kinetic term for $\varphi$
was dropped there, and also a rather atypical coupling
$f(\varphi)\propto \varphi^n$, with $n<0$, was considered.}
discussed constraints on the field-dependent Gauss-Bonnet
couplings with a single scalar field, so as to avoid the
short-scale instabilities or superluminal propagation of scalar
and tensor modes~\footnote{If the existence of a superluminal
propagation is seen only as a transient effect, then the model is
phenomenologically viable. However, a violation of causality or
null energy condition should not be seen as an effect of higher
curvature couplings, at least, for late time cosmology; such
corrections to Einstein's theory are normally suppressed as
compared to the Ricci-scalar term, as well as the scalar
potential, at late times. For a realistic cosmology the quantity
$f(\varphi) H^2$ should be a monotonically decreasing function of
the number of e-folds, $N$, since $f(\varphi) R_{GB}^2 \propto
f(\varphi) H^4$ and $V(\varphi)\propto H^2(\varphi)$.}. These
conditions are nothing but
\begin{equation} 0< c_{\cal R}^2\le 1,
\quad 0< c_T^2 \le 1, \quad 1-\mu >0.
\end{equation}
As long as $\mu< 1$ and $\mu^\prime/\mu<\epsilon$, we also get
$c_{\cal R}^2<1$ and $c_T^2<1$. The short-scale instabilities
observed in~\cite{Calcagni:06ye} corresponding to the value $\mu
\sim 1$, or equivalently $\nu^\prime-2\epsilon\nu \sim 1$, may not
be physical as the condition $\mu \sim 1$ effectively invalidates
the assumptions of linear perturbation theory (see e.g.
~\cite{Hwang&Noh97,Koivisto:2006}). Furthermore, in the limit
$|\mu|\to 1$, other higher order curvature corrections, like cubic
terms in the Riemann tensor, may be relevant. To put this another
way, the Gauss-Bonnet modification of Einstein's theory alone is
possibly not sufficient for describing cosmology in the regime
$\Omega_{GB}\to \Omega_{\rm total}$. For the consistency of the
model under cosmological perturbations, the condition
$|\mu|=|\Omega_{GB}| <2/3$ must hold, in
general~\cite{Amendola:2005,Koivisto:2006}.

Let us first neglect the contribution of matter fields and
consider the special solution (\ref{second-soln}), obtained with a
simple exponential potential. From figure (\ref{tensor-speed1}) we
see that both $c_{\cal R}^2$ and $c_T^2$ temporarily exceed unity,
because of which cosmological perturbations may exhibit a
superluminal scalar or tensor mode. However, this behaviour can be
significantly different in the presence of matter, especially, if
$\varphi$ is allowed to couple non-minimally to matter fields.

The propagation speeds normally depend on the spatial (intrinsic)
curvature of the universe rather than on a specific realization of
the background evolution during the stage of quantum generation of
scalar and tensor modes (or gravity waves). Thus a plausible
explanation for the existence of super-luminal scalar or tensor
modes ($c_{\cal R}^2>1$ or $c_T^2>1$) is that at the initial phase
of inflation the spatial curvature $K$ is non-negligible, while
the interpretation of $c_T^2$ (and $c_{\cal R}^2$) as the
propagation speed is valid only for $K=0$. Additionally, in the
presence of matter, the above result for $c_{\cal R}^2$ does not
quite hold since the scalar modes are naturally coupled to the
matter sector. The result for $c_T^2$ may be applicable as tensor
modes are generally not coupled to matter fields.

From figures (\ref{ct}) and (\ref{cs}) we can see that, for
ans\"atze such as (\ref{f-v-ansatz}), $c_{\cal R}^2$ and $c_T^2$
may become negative, though temporarily, if one allows a larger
slope for the scalar-GB coupling, namely $\alpha\gg \beta$. It is
precisely this last case for which $c_{\cal R}^2$ and $c_T^2$ may
also take values larger than unity at subsequent stages, leading
to superluminal propagation speeds for scalar and/or tensor modes
(see also \cite{Tsujikawa:2006ph,Bonvin:2006} for discussions on a
similar theme). The case where $c_T^2>1$ normally corresponds to
the epoch where the contribution of the coupled Gauss-Bonnet term
becomes significant. However, for smaller values of $\alpha$ and
$\beta$, satisfying $\alpha\lesssim \beta$, both $c_{\cal R}^2$
and $c_T^2$ never become negative and there do not arise any
ghost-like states, though depending upon the initial conditions
the tensors modes may become superluminal, temporarily.

\section{Discussion and Conclusions}

Considering earlier publications on the subject by one of us
(IPN), perhaps it would be useful first to outline the new results
in this paper. Originally~\cite{IPN05d} the model was analysed by
making an appropriate ansatz for the effective potential, namely
\begin{equation}
V(\varphi)+\frac{1}{8} \,f(\varphi) R_{GB}^2=
\Lambda(\varphi)\equiv
H^2(\varphi)\left[3+\epsilon(\varphi)\right],
\end{equation}
which is a kind of gauge choice, and is well motivated by at least
two physical assumptions: firstly, the effective equation of state
$w_{\rm eff} \le 1$ for $\epsilon \ge -3$~\footnote{Only the
stiff-matter can saturate this limit, so $\Lambda(\varphi)=0$, or
equivalently $\epsilon(\varphi)= -3$.}; secondly, this choice
would allow us to find an explicit solution and compute the
primordial spectra more easily. In this work we have relaxed such
a constraint and extended our analysis of the cosmological
solutions of these systems by allowing non-trivial couplings
between the scalar field $\varphi$ and the matter fields. The
cosmological viability of such a generalized theory of
scalar-tensor gravity is fully investigated by placing minimal
constraints on the model parameters and the scalar-matter
couplings.

In summary, we outline the main findings of the paper. We
discussed some astrophysical and cosmological constraints
applicable to a general scalar-tensor gravity models and then
outlined the conditions under which those constraints may be
satisfied for the model under study, with generic choices of the
model parameters. Under the assumptions that the quantity
$\varphi^\prime=\dot{\varphi}/H$ and the time-derivative of the
scalar-GB coupling decrease during inflation being proportional to
some exponential functions of $N\equiv \ln[a(t)]$, namely
$\dot{\varphi}/H\propto e^{\alpha\Z1 N}$ and $\dot{f}H\propto
e^{\alpha\Z2 N}$, the (reconstructed) scalar potential was shown
to take an interesting form $V(\varphi) = H^2(\varphi) (C\Z0 +
C\Z1\, \varphi^2 + C\Z2\, \varphi^{\alpha\Z2/\alpha\Z1})$. Such a
potential, being proportional to the square of the Hubble
parameter, would naturally relax its value as the universe expands
and might have useful implications for the early universe
inflation. With the approximation $\varphi^\prime \approx {\rm
const}$, $V(\varphi)$ was shown to take the form
$V(\varphi)=H^2(\varphi) \left(V\Z0 + V\Z1\,
e^{\beta\varphi}\right)$, for which the effective equation of
state approaches $-1$ without exhibiting any pathological features
at late times.

We have studied the cases where inflaton field $\varphi$ rolls
with a constant velocity ($\varphi^\prime=c$) and/or the universe
experiences a power law inflation, $a(t)\propto a^\alpha$. In
several cases of physical interest, the form of the
(reconstructed) potentials and the scalar-Gauss-Bonnet couplings
are found to be a sum of exponential terms, rather than a single
exponential. We have explicitly shown that, for a spatially flat
metric $K=0$, non-singular inflationary solutions are possible for
a wide range of scalar-GB couplings. We have found explicit
cosmological solutions by imposing minimal restrictions on model
parameters; for example, we allowed the scalar potential to take
the simplest exponential form, but without specifying the coupling
$f(\varphi)$. The model was also shown to admit an epoch of
inflation for a vanishing potential, $V(\varphi)=0$. This should
not come as a surprise since we have already known from
Starobinsky's work~\cite{Starobinsky:80a} that inflation can
easily be triggered by the 4D anomaly driven $R^2$-terms. However,
in this case, although the spectral indices of cosmological
perturbations may be brought close to observationally supported
values, such as $n\Z{s}\approx 0.96$, it is generally difficult to
obtain an inflationary solution supporting a small
tensor-to-scalar ratio, $r\ll 0.5$.

We have shown (both analytically and numerically) that getting
only a small deviation from $w=-1$ in our model necessarily
implies a non-trivial scalar-GB coupling. The question of course
is how natural is the existence of the phase $w <-1$ and the phase
supporting matter to dark energy dominance. Depending upon the
initial values of the model parameters, a momentary crossing of
the EoS $w=-1$ may be realized in various contexts. We have
explained under what conditions $w+1$ can be negative, given the
required period of matter-dominance is maintained before realizing
the currently observed acceleration of the universe. We have also
established rather systematically that in theories of
scalar-tensor gravity with the coupled Gauss-Bonnet term, the
quantity $w+1$ can be negative only momentarily, while approaching
the limit $w=-1$ from above. Hence indicating that none of the
issues inherent with a super-luminal propagation or a violation of
unitarity will be applicable for a late time cosmology. It would
be of interest to extend our analysis to the $K=\pm 1$ cases.

Our analysis is not conclusive in answering whether the model
under consideration is viable for describing our universe from the
epoch of nucleosynthesis until today without any violation of
unitarity, for instance, due to an imaginary propagation speed for
the tensor or scalar modes. Such violations imply the existence of
some sort of ghost states, or superluminal propagation, implying a
break down of the weak energy condition or causality.
Nevertheless, such pathological features of the model observed for
simplest choices of the potential, e.g. $V(\varphi)\propto
e^{-\,\beta\varphi}$ and $f(\varphi)\propto e^{\,\alpha\varphi}$,
may be avoided by introducing scalar-matter couplings and/or by
suitably choosing values of the slope parameters $\alpha$ and
$\beta$ that are phenomenologically viable. Extension to a more
general situation with $V(\varphi)\to V(\varphi,\sigma)$ where the
latter is a sum of exponential terms can be made rigourously by
considering another scalar field $\sigma$. This will appear
elsewhere.

\section*{Acknowledgements}

We wish to thank David Mota, David Wiltshire and Shinji Tsujikawa
for constructive remarks on the draft, and Sergei Odintsov,
Shin'ichi Nojiri for helpful correspondence. The work of IPN is
supported by the Foundation for Research, Science and Technology
(NZ) under Research Grant No. E5229, and of BML by the Marsden
Fund of the Royal Society of New Zealand.

\appendix
\section{Appendix:}
\renewcommand{\theequation}{A.\arabic{equation}}
\setcounter{equation}{0}

In the appendices below we will present some technical details
relevant to solving the field equations, along with a linear
stability analysis of a critical solution.

\subsection{Basic equations}

We find it convenient to define
\begin{equation}
f_{,\,\varphi} H^2(\varphi)  \equiv u(\varphi), \quad
\frac{\dot{\varphi}}{H}=\varphi^\prime\equiv x,
\end{equation}
where a prime denotes the derivative with respect to the number of
e-folds, $N=\int H\, {d t}=\ln (a(t)/a_0)$. A simple calculation
shows that
\begin{eqnarray}
\ddot{f}=(u x)^\prime -\epsilon (ux),\quad
\ddot{\varphi}=(x^\prime+\epsilon x) H^2, \quad
\frac{1}{H^2}\,V_{,\,\varphi}= \frac{y^\prime+2\epsilon}{x}.
\end{eqnarray}
Equations (\ref{GB1})-(\ref{GB3}) may be expressed as the
following set of autonomous equations:
\begin{eqnarray} && 0=
-3+\frac{\gamma\kappa^2}{2}\, x^2 +
\kappa^2 y+ 3 \kappa^2 x u+ 3\Omega_m+ 3\Omega_r,\label{newA4}\\
&& 0= 2\varepsilon+3- \kappa^2(x u)^\prime-\kappa^2
(\varepsilon+2)x u +\frac{\gamma\kappa^2}{2}\,x^2 -\kappa^2 y+3
w_m \Omega_m+ \Omega_r,\label{newA5}\\
&&0= \gamma\kappa^2 x x^\prime+\kappa^2(\varepsilon+3)\gamma x^2
+\kappa^2 (y^\prime+2\varepsilon y) + 3\kappa^2 (1+\varepsilon) x
u-3\eta Q \Omega_m x, \label{newA6}
\end{eqnarray}
where $\eta=1, -2$ and $0$ respectively for ordinary matter
($w_m=0$), stiff matter ($w_m=1$) and radiation ($w_r=1/3$). The
continuity equation (\ref{conservation-eq}) may be given by
\begin{eqnarray}
&&0=\Omega^\prime_m + 2\epsilon \Omega_m +3(1+w_m)\Omega_m+3\eta
Q \Omega_m x,\\
&& 0= \Omega^\prime_r+ 2(\epsilon+2) \Omega_r.
\end{eqnarray}
In the case of minimal coupling, so $Q=0$, we $\rho_b\propto
e^{-3(1+w_b) N}$. This last expression implies that the energy
density due to radiation ($w_b=1/3$) decreases faster (by a factor
of $e^{N}$) than due to the ordinary matter ($w_b=0$).

\subsection{A special solution}

As a canonical choice, consider the potential $V(\varphi)$ and the
coupling $f(\varphi)$ as the simplest exponential functions of the
field:
\begin{equation}\label{f-and-V}
V(\varphi)= V_1\, {\rm e}^{m\varphi/\varphi_0}, \quad
f(\varphi)=f_1\, {\rm e}^{n\varphi/\varphi_0}.
\end{equation}
These choices may be motivated in the heterotic string theory as
the first term of the perturbative string
expansion\cite{Antoniadis93a}. One may solve the
equations~(\ref{GB1})-(\ref{GB3}) by expressing $V(\varphi)$ and
$f(\varphi)$ as in~(\ref{f-and-V}), with $n=-m=2$, and dropping
the effects of matter fields. The simplest (and also perhaps the
modest) way of solving field equations is to make a particular
ansatz for the scale factor, as well as for the scalar field,
namely
\begin{eqnarray}
&& a = a_0 \left(\frac{t}{t_0}\right)^{h}, \quad \varphi
=\varphi_0 \ln \frac{t}{t_1} \quad (h >0,~
t>0),\label{asymptotes1}
\\
&& a = a_0 \left(\frac{t_\infty-t}{t_0}\right)^{h}, \quad \varphi
= \varphi_0 \ln \frac{t_\infty-t}{t_1} \quad (h<0,~ t<0
~\mbox{or}~0<t<t_\infty).\label{asymptotes2}
\end{eqnarray}
We can then reduce the field equations in the following form:
\begin{eqnarray}
0&=& -\frac{3h^2}{\kappa^2} +\frac{\gamma\varphi_0^2}{2} +V_1
t_1^2+\frac{6 f_1 h^3}{t_1^2},
\\
0&=& \frac{h(3h-2)}{\kappa^2} +\frac{\gamma\varphi_0^2}{2} -V_1
t_1^2 -\frac{2 f_1 h^3 (2h-1)}{t_1^2},
\\
0&=& \gamma(1-3h)\varphi_0^2 +2 V_1 t_1^2 -\frac{6 f_1 h^3
(h-1)}{t_1^2}.
\end{eqnarray}
By solving any two of these equations, we find
\begin{eqnarray}
V_1 t_1^2 &=& \frac{(5h-1)\gamma \varphi_0^2 \kappa^2+6h^2
(h-1)}{\kappa^2(1+h)},\nonumber
\\
\frac{2 f_1 h^2}{t_1^2} &=& \frac{2h-\gamma\varphi_0^2
\kappa^2}{\kappa^2(1+h)}, \quad \varphi_0=\varphi_0,
\end{eqnarray}
for $h\neq -1$, while
\begin{equation}
\varphi_0^2=-\frac{2}{\gamma\kappa^2}, \quad V_1
t_1^2=\frac{4}{\kappa^2}+ \frac{6 f_0}{t_1^2}, \quad f_1=f_1,
\end{equation}
for $h=-1$. By fixing the values of $V_0$ and $t_1$, one can fix
the value of $\varphi_0$, in terms of the expansion parameter $h$,
which itself needs to be determined by observation. The equation
of state parameter can be written as
\begin{equation}
w= -1+\frac{2}{3 h}.
\end{equation}
Of course, $w<-1$ is possible for $h<0$. In this case the Hubble
rate would increase with proper time $t$, since
$H=-{h}/{(t_\infty-t)}$. The observational results, coming from
Hubble parameter measurements using Hubble space telescope and
luminosity measurements of Type Ia supernovae, appear to indicate
the value~\cite{Wang04py}
\begin{equation}
-1.62 < w < -0.74 \nonumber
\end{equation}
at $95\%$ confidence level. One may think of this value of $w$ as
two separate regimes:
\begin{equation}
-1.62 < w < -1, \qquad  -1 < w <-0.74 \nonumber \end{equation}
which, for the above critical solution, require that $ h < -1.0753
$ or $ 2.5641< h < \infty$. That means, even if $\gamma>0$, as is
the case for a canonical $\varphi$, then the Gauss-Bonnet gravity
could support both the effective phantom and the quintessence
phase of the late universe. This was also one of the main points
of the Gauss-Bonnet dark energy proposal made by Nojiri {\it et
al} in \cite{Nojiri05b}, where the Gauss-Bonnet potential
$V_{GB}\equiv f(\varphi)\,{\cal R}_{GB}^2$ is important, although
it contributes to the effective action only subdominantly.

Although the special solution above avoids the initial singularity
at $t=0$, it develops a big-rip type singularity at an asymptotic
future $t=t_\infty$. Such solutions are typically only the fixed
point, or critical solutions, which in general are not applicable
once the matter degrees of freedom are taken into account, both
for minimal and non-minimal couplings of $\varphi$ with matter.
Nevertheless, it may be interesting to check stability of such
solutions under linear (homogeneous) perturbations, when
$V(\varphi)\ne 0$, which was not done in~\cite{Nojiri05b}.

\subsection{A linear stability analysis}

Our analysis in this subsection provides some new insights into
the nature of instability of the critical solution under a linear
(homogeneous) perturbation; the earlier analysis
in~\cite{Nojiri05b} would be a special case of our treatment.

With the ansatz (\ref{asymptotes1})-(\ref{asymptotes2}), and the
choice (\ref{f-and-V}) and $m=-n=-2$, the system of autonomous
equations, in the absence of matter fields, is given by
\begin{eqnarray}
\frac{dx}{dN}&=&\frac{\gamma^2\varphi_0 x^3-\gamma x (2u x^2-13 u
x\varphi_0+6\varphi_0)+2(2y-3ux)-6u(\varphi_0 -2x u\varphi_0+ x^2
u)}
{\varphi_0(2\gamma-2\gamma x u+3 u^2)},\nonumber \\ \\
\frac{du}{dN}&=& \frac{2u\left(-\gamma^2 x^2
\varphi_0+u(2y-3(\varphi_0-x)u)+ 2\gamma (x-2\varphi_0
u)\right)}{\varphi_0(2\gamma-2\gamma x u + 3 u^2)},
\end{eqnarray}
where $y\equiv V/H^2$. For the following special (critical)
solution
\begin{equation} x=x_0=\frac{\varphi_0}{h}, \qquad y=y_0=
\frac{V_1 t_1^2}{h^2}, \qquad u=u_0= \frac{2 f_1 h^2}{\varphi_0
t_1^2},
\end{equation}
the right hand sides of the above equations are trivially
satisfied, implying that $x^\prime=y^\prime=z^\prime=0$.
\begin{figure}[ht]
\begin{center}
\epsfig{figure=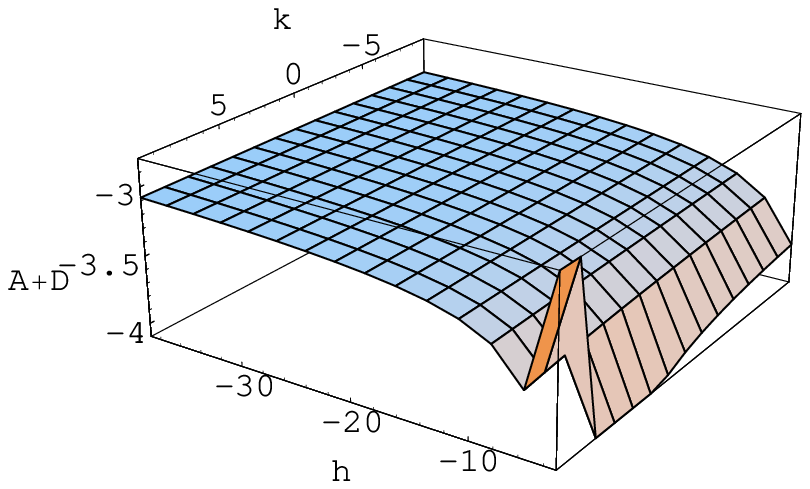,height=2.2in,width=2.8in} \hskip0.3in
\epsfig{figure=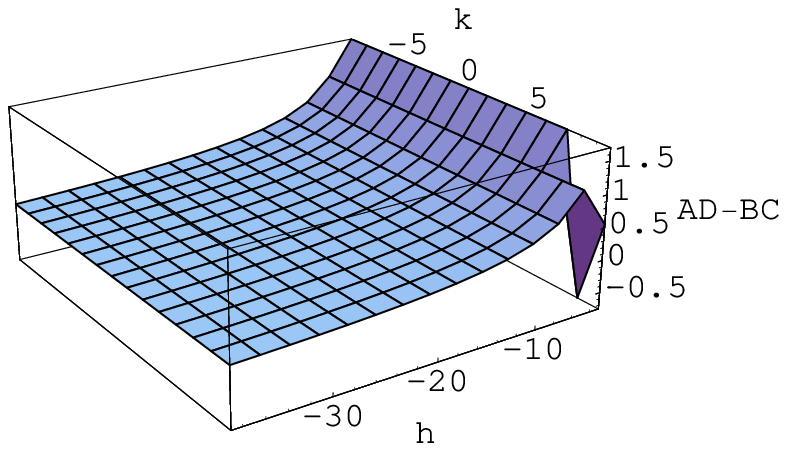,height=2.2in,width=2.8in}
\end{center}
\caption{Only for negative eigenvalues, satisfying the conditions
$A+D<0$ and $AD-BC>0$ the linear perturbations become small and
the system may be stable.} \label{eigen-val}
\end{figure}

Let us consider the following (homogeneous) perturbation around
the critical solution (\ref{asymptotes1}) or (\ref{asymptotes2}):
\begin{equation}
x=x_0 +\delta x , \qquad u= u_0+ \delta u.
\end{equation}
Since the functional form of the potential is already specified,
the variation of $y$ is encoded into the variation of $\varphi$.
For stability of the solution, under a small perturbation about
the critical or fixed point solution, requires
\begin{equation}
A+D< 0, \qquad AD- BC >0,
\end{equation}
where $A,B, C, D$ are the eigenvalues of a $2\times 2$ matric, as
given by
\begin{equation}
\frac{d}{dN}\left(
\begin{array}{c}
\delta x\\ \delta z \end{array}\right)= M\left(\begin{array}{c}
\delta x \\ \delta z\end{array}\right), \qquad
M=\left(\begin{array}{c} A \qquad B \\
C \qquad D\end{array}\right).
\end{equation}
In the following, we define $V_1 t_1^2\equiv k$. We find
\begin{equation}
A =-\frac{\tilde{A}}{h E}, \quad B=-\frac{\tilde{B}}{\gamma E},
\quad C=-\frac{\gamma \,\tilde{C}}{3h^2(3h^2-3h-2k) E}, \quad
D=-\frac{\tilde{D}}{h
 E}
\end{equation}
where
\begin{eqnarray}
\tilde{A}&\equiv& 18h^4(h-1)(9h^2-4h+1) +3k (1-24 h) h^2+k^2(132
h^3-22 h^2-7h+1), \nonumber \\ \\
\tilde{B} &\equiv& 3 (3h^2-3h-2k) \left( 6 h^2(3h^2-2h+1)-k(22h^2-13 h+1)\right),\\
\tilde{C} &\equiv & \left(36 h^5(3h^2-1)- 2k h^2(72 h^3+42 h^2-12
h(3h+2)+6) + 2k^2(28h^2-12 h+1)\right)\nonumber\\ &{}& \times\,
(9h^3-3h^2-6hk +k),\\
\tilde{D} &=&36 h^4(3h-1)(h-1)^2+2 k^2 (24 h^3-68 h^2+17h-1)- 4 k
h^2(36 h^3 -93 h^2+51 h-6).\nonumber \\ \\
{E} & =& 18 h^4 (5h^2-4 h+1)+k^2(60 h^2-16 h+1)- 6k h^2 (25 h^2+
17 h-2).
\end{eqnarray}
We find the result in~\cite{Nojiri05b} by taking
$k=0$~\footnote{The results here also correct errors/typos that
appeared in the appendix of~\cite{Nojiri05b}.}. In the case
$\gamma>0$ and $h<0$, all eigenvalues are negative, except when
$|h|$ is small, $|h|\lesssim 2$, or when the parameter $k$ takes a
large value (cf figure \ref{eigen-val}). While, for $\gamma<0$,
two of them ($B$ and $C$) take positive values, leading to a
classical instability of the critical solution. Such a system is
normally unstable also under inhomogeneous (cosmological)
perturbations. In particular, for a large and positive potential,
so that $k\gg 0$, only the $h>0$ solution can be stable.

\subsection{Fixed dilaton/modulus}

It is quite plausible that after inflation the field $\varphi$
would remain almost frozen for a long time, so $x\simeq 0$,
leading to an universe filled dominantly with radiation and highly
relativistic or stiff matter. Such a phase of the universe is
described by the solution
\begin{equation}
 \Omega_m^{w=1}=\frac{1}{1+\Omega_1}, \quad
\Omega_r=\frac{\rho_1\,e^{2N}}{1+\Omega_1}, \quad
\epsilon=\frac{-3-2\rho_1\,e^{2N}}{1+\Omega_1},\quad y=\frac{3
\rho_2\,e^{6N}}{1+\Omega_1}
\end{equation}
where $\Omega_1 \equiv \rho_1\,e^{2N} + \rho_2\,e^{6N}$ and
$\rho_i$ are integration constants. One takes $\rho_i>0$, so that
$\Omega_r>0$ and $V(\phi)\approx \Lambda$. The density fraction
for a stiff fluid, $\Omega_m^{w=1}$, decreases quickly as the
universe further expands, $N\gg 0$, thereby creating a vast amount
of thermal radiation.

While, at late-times, $\varphi$ is again rolling slowly, such that
$x\approx 0$. The cosmic expansion is then characterized by
\begin{equation}
\Omega_m^{w=0}=\frac{1}{1+\Omega_2}, \quad \Omega_r=\frac{\delta_1
e^{- N}}{1+\Omega_2}, \quad \epsilon=\frac{- 3- 4
\delta_1\,e^{-N}}{2+2\Omega_2},\quad
y=\frac{3\delta_1\,e^{3N}}{1+\Omega_2}
\end{equation}
where $\Omega_2 \equiv \delta_1\,e^{- N}+ \delta_2\,e^{3N}$ and
$\delta_i$ are integration constants. Both $\Omega_m$ and
$\Omega_r$ decrease with the expansion of the universe, but unlike
a universe dominated by highly relativistic fluids or
stiff-matter, the radiation energy density decreases now faster as
compared to the matter energy density, as it is red-shifted away
by a factor of $e^{-N}$.

\end{document}